\documentclass[lettersize,journal]{IEEEtran}
\usepackage{amsmath,amsfonts}
\usepackage{algorithmic}
\usepackage{algorithm}
\usepackage{array}
\usepackage[caption=false,font=normalsize,labelfont=sf,textfont=sf]{subfig}
\usepackage{textcomp}
\usepackage{stfloats}
\usepackage{url}
\usepackage{verbatim}
\usepackage{graphicx}
\usepackage{cite}
\usepackage{color}
\usepackage{amssymb}
\usepackage{multirow}

\usepackage{booktabs} 
\usepackage{flushend}
\usepackage{hyperref} 
\hypersetup{
  colorlinks=true,
  citecolor=blue,
  linkcolor=red,
  urlcolor=green}

\newcommand{\cov}{\boldsymbol{\Theta}}
\newcommand{\bLambda}{\boldsymbol{\Lambda}}
\newcommand{\blambda}{\boldsymbol{\lambda}}
\newcommand{\bPi}{\boldsymbol{\Pi}}

\newcommand{\bSigma}{\boldsymbol{\Sigma}}

\newcommand{\rmR}{\mathrm{R}}
\newcommand{\rmT}{\mathrm{T}}
\newcommand{\rmd}{\mathrm{d}}

\newcommand{\rmA}{\mathrm{A}}

\newcommand{\bh}{\mathbf{h}}

\newcommand{\bb}{\mathbf{b}}
\newcommand{\bB}{\mathbf{B}}
\newcommand{\by}{\mathbf{y}}

\newcommand{\bn}{\mathbf{n}}
\newcommand{\bg}{\mathbf{g}}
\newcommand{\bA}{\mathbf{A}}
\newcommand{\bH}{\mathbf{H}}
\newcommand{\bU}{\mathbf{U}}
\newcommand{\bu}{\mathbf{u}}
\newcommand{\bQ}{\mathbf{Q}}
\newcommand{\bI}{\mathbf{I}}
\newcommand{\bw}{\mathbf{w}}
\newcommand{\bx}{\mathbf{x}}
\newcommand{\bM}{\mathbf{M}}

\newcommand{\sfC}{\mathrm{C}}

\newcommand{\CN}{\mathcal{CN}}
\newcommand{\calU}{\mathcal{U}}

\newcommand{\bbC}{\mathbb{C}}

\newcommand{\bbE}{\mathbb{E}}

\newcommand{\ub}{\mathsf{ub}}
\newcommand{\sparse}{\mathsf{SS}}
\newcommand{\RS}{\mathsf{DS}}

\newcommand{\ctrans}{\mathsf{H}}
\newcommand{\half}{\frac{1}{2}}
\newcommand{\halfH}{\frac{\mathsf{H}}{2}}
\newcommand{\rank}{\mathrm{rank}}
\newcommand{\diag}{\mathrm{diag}}
\newcommand{\Tr}{\mathsf{Tr}}
\newcommand{\opt}{\mathsf{opt}}
\newcommand{\threshold}{\mathsf{th}}
\newcommand{\Prob}{\mathsf{Pr}}
\newcommand{\out}{\mathsf{out}}
\newcommand{\app}{\mathsf{app}}

\newcommand{\angwid}{\varrho}

\newtheorem{proposition}{Proposition}%{\textbf{Proposition}}
\newtheorem{remark}{Remark}%{\textbf{Remark}}
\newtheorem{theorem}{Theorem}%{\textbf{Remark}}
\newtheorem{lemma}{Lemma}

\begin{document}

\title{Mid-Band Extra Large-Scale MIMO System: Channel Modeling and Performance Analysis}

\author{
Jiachen Tian,~\IEEEmembership{Graduate Student Member,~IEEE,}
Yu Han,~\IEEEmembership{Member,~IEEE,}
Xiao Li,~\IEEEmembership{Member,~IEEE,}\\
Shi Jin,~\IEEEmembership{Fellow,~IEEE,}
and Chao-Kai Wen,~\IEEEmembership{Fellow,~IEEE}

\thanks{J. Tian, Y. Han, X. Li and S. Jin are with the National Mobile Communication Research Laboratory, Southeast University, Nanjing 210096, China (email: tianjiachen@seu.edu.cn; hanyu@seu.edu.cn; li\_xiao@seu.edu.cn; jinshi@seu.edu.cn).}
\thanks{C.-K. Wen is with the Institute of Communications Engineering, National Sun Yat-sen University, Kaohsiung 80424, Taiwan (e-mail: chaokai.wen@mail.nsysu.edu.tw).}
}

\maketitle

\begin{abstract}
In pursuit of enhanced quality of service and higher transmission rates, communication within the mid-band spectrum, such as bands in the 6-15 GHz range, combined with extra large-scale multiple-input multiple-output (XL-MIMO), is considered a potential enabler for future communication systems. However, the characteristics introduced by mid-band XL-MIMO systems pose challenges for channel modeling and performance analysis. In this paper, we first analyze the potential characteristics of mid-band MIMO channels. Then, an analytical channel model incorporating novel channel characteristics is proposed, based on a review of classical analytical channel models. This model is convenient for theoretical analysis and compatible with other analytical channel models. Subsequently, based on the proposed channel model, we analyze key metrics of wireless communication, including the ergodic spectral efficiency (SE) and outage probability (OP) of MIMO maximal-ratio combining systems. Specifically, we derive closed-form approximations and performance bounds for two typical scenarios, aiming to illustrate the influence of mid-band XL-MIMO systems. Finally, comparisons between systems under different practical configurations are carried out through simulations. The theoretical analysis and simulations demonstrate that mid-band XL-MIMO systems excel in SE and OP due to the increased array elements, moderate large-scale fading, and enlarged transmission bandwidth.
\end{abstract}


\section{Introduction}

\IEEEPARstart{T}{he} fifth-generation (5G) wireless communication system, operational since 2020, has significantly advanced mobile technology. As the demand for higher data rates and improved quality of service grows, research and development have expanded to include the sixth-generation (6G) wireless systems, which are under exploration in both academic and industrial sectors \cite{MatthaiouCM}. A pivotal innovation for these future mobile systems is extra-large-scale massive MIMO (XL-MIMO) \cite{HanIOTJ}, which boosts transmission rates and supports communication with multiple users simultaneously. Additionally, spectrum usage continues to be a critical issue. The Sub-6 GHz band, while effective in 5G for achieving high capacity and reliability, now faces bandwidth limitations. Conversely, millimeter-wave (mmW) frequency bands, despite their severe path loss issues, offer vast frequency resources crucial for expanding capacity in the 5G era. Meanwhile, the mid-band spectrum, specifically the 6-24 GHz range, is emerging as a strategic choice for international mobile telecommunications (IMT) services in 6G, offering substantial capacity and competitive coverage \cite{nokia}.

Recent developments from the World Radiocommunication Conference 2023 (WRC-23) \cite{WRC23} are notable, with the allocation of up to 700 MHz of spectrum in the upper 6 GHz band and more than 2 GHz in the 7-15 GHz range for exclusive and/or shared use.\footnote{WRC-23 addressed mid-band spectrum issues, identifying the upper 6 GHz band (6.425-7.125 GHz) for IMT across Europe, the Middle East, Africa, and selected countries in the Americas and Asia. It also initiated a new IMT/6G study for WRC-27, focusing on frequency bands in the 7-15 GHz range.} This so-called upper mid-band spectrum is anticipated to be the primary frequency for IMT services in the 6G era. The forthcoming 6G standard is expected to utilize a wide range of mid-bands, including the Sub-6 GHz and upper mid-band. However, the deployment of mid-band\footnote{In this paper, 'mid-band spectrum' generally refers to the frequency bands in the 1-24 GHz range, encompassing both the Sub-6 GHz band and upper mid-band.} XL-MIMO systems introduces new challenges in channel modeling and system design, requiring innovative solutions to fully exploit their potential and meet the evolving needs of future wireless communications.

\subsection{Related Works}

Research has illustrated the feasibility of the upper mid-band spectrum \cite{U6G,U6G1}. Considering the wideband characteristics and overlap with satellite spectrum, spectrum management strategies and interference suppression schemes have been proposed \cite{spectrum}. Beyond intuitive analysis, ray-tracing (RT) based simulations are carried out to reveal the potential of the upper mid-band in coverage and throughput \cite{RT}. Meanwhile, practical measurement campaigns of upper mid-band channel characteristics hold equal significance, on which research has primarily focused so far. The authors in \cite{MiaoJSAC} measured and compared the channel characteristics of the 3.3 GHz, 6.5 GHz, 15 GHz, and 28 GHz bands under two typical scenarios, and revealed the potential of the upper mid-band spectrum in network coverage and transmission reliability, which validates the effectiveness of the upper mid-band spectrum and provides guidance for the design of transmission strategies. In \cite{TSR}, detailed comparisons of indoor channel characteristics were presented between emerging frequency bands belonging to frequency range1 (FR1) and FR3, respectively. Regarding scattering characteristics, extensive research and measurement efforts have been directed towards the phenomenon of dense multipath components (DMC) within the upper mid-band spectrum \cite{DMC1,DMC2}, in which the channel characteristics in the angular domain and frequency domain are modeled from a statistical perspective.

Leveraging the upper mid-band spectrum enables the integration of more elements within an array compared with the Sub-6 GHz frequency band. Therefore, the upper mid-band spectrum is expected to be integrated synergistically with XL-MIMO systems \cite{TianWCL,LuICC,YangTVT,LDMA,NFCor,XLmeasure}. As an enhanced regime of massive MIMO, XL-MIMO is anticipated to confer advantages, such as augmenting transmission streams and enabling massive user access. Specifically, the authors in \cite{LuICC} investigated the scaling of the signal-to-noise ratio (SNR) when the number of antennas tends to infinity. The ergodic spectral efficiency (SE) under linear receivers of XL-MIMO systems is analyzed in \cite{YangTVT}. Additionally, the near-field property is regarded as an important characteristic of XL-MIMO systems. Based on near-field channel conditions, the authors in \cite{LDMA} proposed a location division multiple access (LDMA) scheme and illustrated the enhancement in SE. Considering the spatial correlation characteristics, the authors in \cite{NFCor} modeled the near-field correlation characteristics through a modified one-ring model. 
From a practical measurement perspective, the authors in \cite{XLmeasure} conducted measurement campaigns of XL-MIMO channels and proposed empirical models of key channel characteristics, validating the near-field non-stationary property. 

\subsection{Key Problems}

The mid-band spectrum, spanning from the lower portion near the Sub-6 GHz band to the upper portion near the mmW band, presents unique challenges and opportunities. Although existing state-of-the-art studies have focused on mid-band wireless communication aspects such as feasibility, measurement, RT simulations, and spectrum strategies, channel modeling and performance analysis for mid-band XL-MIMO systems still emerge as critical concerns, especially from a theoretical perspective. Such channel modeling requires high analyzability while maintaining concise expressions to guide the design of transmission strategies and explore potential performance improvements in future wireless communication systems.
Specifically, the following problems exist:
 
{\bf P1:} 
An analytical channel model that encompasses the main characteristics of XL-MIMO systems under a wide mid-band spectrum is lacking. This model should exhibit strong compatibility and convenience for performance evaluation. 

{\bf P2:} 
The potential of mid-band XL-MIMO systems remains unclear. Necessary performance analysis and comparisons with systems under different configurations, such as frequency bands, should be carried out.

\subsection{Contributions}
In this paper, the problems above are tackled through the following contributions:
\begin{itemize}

\item \textit{Integrating measurement insights and analytical approaches for mid-band XL-MIMO channel modeling}.
Based on existing measurement results, the characteristics of the mid-band channel are summarized. Specifically, given the integration with a large-scale array aperture, the channel modeling is expected to account for near-field non-stationary scenarios. Meanwhile, sparsity and spatial correlation are also considered inherent properties due to the broad range of the mid-band spectrum and its multiple application scenarios. By reviewing typical analytical channel models from the perspectives of correlation and propagation, we pave the way for channel modeling of the mid-band XL-MIMO channel.

\item \textit{A pervasive analytical channel model and corresponding compatibility analysis}.
To better explain the proposed analytical channel model, we illustrate the scenario of mid-band XL-MIMO systems.
An analytical channel model is proposed as the solution to {\bf P1}, which embraces different frequency bands and scenarios through flexible and adjustable parameter configurations, as well as is compatible with other analytical channel models.
Specifically, the compatibilities of the proposed channel model with distance, scenarios, and other analytical channel models are analyzed.

\item \textit{Revealing the potential of mid-band XL-MIMO systems through analysis and simulations of key performance metrics}.
Based on the proposed model, key performance metrics including ergodic SE and outage probability (OP) of MIMO maximal-ratio combining (MRC) systems are analyzed. Specifically, closed-form approximations and performance bounds are derived for the two typical systems.
On this foundation, we focus on spatial correlation matrices and delve into the characteristics of their eigenvalues.
Meanwhile, extensive numerical results are provided based on practical system configurations to compare the performance of systems under different frequency bands. The theoretical and simulation results reveal the advantages of mid-band XL-MIMO systems from the perspective of increased array elements, near-field spatial correlation characteristics, and enlarged transmission bandwidth, which provides a solution to {\bf P2}.
\end{itemize}

The rest of this paper is organized as follows.
Section II provides preliminaries of mid-band channel modeling.
The proposed analytical channel model and corresponding analysis are illustrated in Section III.
In Section IV, key performance metrics are analyzed and discussed based on the proposed channel model.
The simulation results based on practical configurations are provided in Section V.
Finally, conclusions are given in Section VI.

\textit{Notations}--Vectors and matrices are denoted by bold lowercase and uppercase letters, respectively.
The superscripts $(\cdot)^{\top}$ and $(\cdot)^{\mathsf{H}}$ represent the transpose and conjugate transpose, respectively.
The expectation is denoted by $\bbE \{ \cdot \}$; $\otimes$ and $\odot$ represent the Kronecker product and the Hadamard product, respectively;
$\mathrm{diag}(\cdot)$ is a diagonal matrix;
$\Tr(\bM)$, $\rank(\bM)$ and $\blambda(\bM)$ are the trace, the rank, and a vector containing eigenvalues of matrix $\bM$, respectively.
The element in the $m$-th row and $n$-th column of matrix $\bM$ is denoted as $[\bM]_{m,n}$.
A circular symmetric complex Gaussian random variable with mean $m$ and variance $\iota^2$ is denoted as $\mathcal{CN}(m,\iota^2)$, and $\mathcal{U}(a,b)$ is a uniform distribution.

\section{Preliminaries of Mid-Band Channel Modeling}
In this section, we first summarize the potential characteristics of the mid-band XL-MIMO channel, aiming to elucidate the requirements for channel modeling of mid-band XL-MIMO systems. Afterward, we review five typical analytical MIMO channel models and compare their advantages and disadvantages in characterizing the mid-band XL-MIMO channel.

\begin{table*}[!t]
	\caption{Comparisons of different analytical channel models\label{tab:model_cmp}}
	\centering
	\begin{tabular}{cccccc}%{ccccc}%|c|c|c| |c|c|c|c|
	\toprule
	\multirow{2}*{Model} & \multirow{2}*{Category} & Propagation & Spatial & Near-field & Parameterized through \\
    ~ & ~ & mechanism & correlation &  properties &  different measurements\\
	\midrule
	%\hline
    Finite-dimensional model \cite{finite} & PBSM & Yes & No & No & Limited\\
    Virtual channel representation \cite{VCR} & PBSM & Yes & No & No & Limited\\
	Kronecker model \cite{Kronecker} & CBSM & No & Yes & No & Limited\\
	Double-scattering model \cite{double} & CBSM & No & Yes & No & Limited\\
    Weichselberger model \cite{Weichselberger} & CBSM & No & Yes & No & Limited\\
	\bottomrule
	\end{tabular}
\end{table*}

\subsection{Potential Characteristics of Mid-Band XL-MIMO Channels}
\label{sec:XLMIMOfeature}

\begin{itemize}
\item \textit{Near-Field Effect}:
    The utilization of an extra large-scale array extends the range of the near field, making it more likely for UEs and clusters to appear in the near field of the array. Traditional models, which assume planar wavefronts, may not accurately reflect the real propagation environment. Therefore, it is warranted to incorporate a near-field model that considers spherical wavefronts while ensuring compatibility with far-field situations.

\item \textit{Non-Stationarities}:
    Theoretically, received signal power can be modeled as inversely proportional to the distances between the Tx and Rx array elements. Therefore, the variation of large-scale fading along array elements cannot be ignored when an extra large-scale array is employed, leading to noticeable unevenness in the received power of array elements. Measurement results in \cite{XLmeasure,nonstationary} have validated the fluctuation of signal power received by different array elements. 

\item \textit{Cluster-Level Sparsity}:
    Measurement results indicate that clusters under mid-band channel conditions are distinguishable \cite{MiaoJSAC,U6Gsparse}, especially in the higher portions of the mid-band spectrum. Thus, the assumption of a rich-scattering environment, such as the i.i.d. Rayleigh channel condition, is challenging to uphold. From another perspective, ensuring compatibility with sparsity is also important for different scenarios, including indoor and outdoor scenarios at short ranges.

\item \textit{Spatial Correlation}:
    Spatial correlation is a critical channel characteristic that significantly influences system performance, closely associated with the angular distribution under different frequency bands. Meanwhile, in the lower parts of the mid-band, the channel tends to exhibit rich scattering characteristics, whilst sparse scattering characteristics tend to emerge in the higher parts of the mid-band. Therefore, from a statistical perspective, it is also preferable to describe the random scattering characteristics of the mid-band XL-MIMO channel through spatial correlation matrices.
\end{itemize}

\subsection{Review of Analytical MIMO Channel Models}

Considering the aforementioned characteristics, the development of novel channel models assumes significant importance. Generally, channel models can be categorized into two main streams: physical models and analytical models \cite{channelsurvey}. Physical models include deterministic channel models and the geometry-based stochastic channel model (GBSM). Analytical models are further divided into the correlation-based stochastic channel model (CBSM) and the propagation-based stochastic channel model (PBSM). 

Although physical models offer high accuracy, they are also marked by extremely high complexity, which is not conducive to theoretical performance analysis.  
On the other hand, analytical channel models, including CBSMs and PBSMs, with their simpler but critical structures, prove advantageous for the design of transmission strategies and capacity analysis. PBSMs characterize the channel by parameters such as delay, angle of arrival (AoA), and complex gain through a simplified representation, while CBSMs describe the channel from the perspective of temporal and spatial correlation matrices.
Reviewing the extensive state-of-the-art analytical channel models \cite{finite,VCR,Kronecker,double,Weichselberger}, the pros and cons of characterizing mid-band MIMO channels are listed in Table \ref{tab:model_cmp}, with specific descriptions of each model found in \cite{channelsurvey,CJE}.

The classic analytical channel models have been highly successful in characterizing both Sub-6 GHz and mmW channels. However, as shown in Table \ref{tab:model_cmp}, classic analytical channel models focus either on the propagation mechanism or the spatial correlation characteristics.
Meanwhile, these models fail to incorporate emerging channel characteristics of mid-band XL-MIMO channels, such as near-field properties and spatial non-stationarities, and cannot flexibly balance the sparsity and scattering characteristics. Furthermore, they cannot be parameterized flexibly based on different measurement results. 
Therefore, it is necessary to develop an analytical channel model for mid-band XL-MIMO systems.

\begin{figure}[!t]
    \centering
    \includegraphics[scale=0.55]{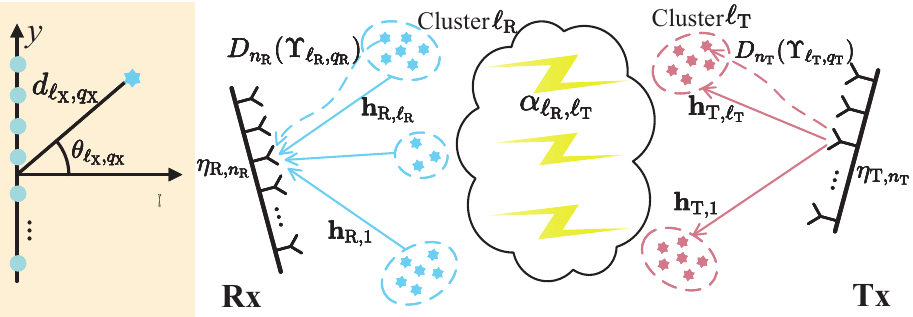}%0.55
    \caption{Illustration of mid-band XL-MIMO system.}
    \label{fig:scenario}
\end{figure}

\section{Channel Modeling for Mid-Band XL-MIMO Systems}
To address the problem {\bf P1}, the proposed analytical channel model is detailed in this section. The discussion begins with the system configuration, describing the scenario of mid-band XL-MIMO systems through the lens of a cluster-based channel model. Following this, the proposed analytical channel model, abstracted from the scenario, is elaborated upon. Subsequent discussions focus on the model's compatibility with various distances, scenarios, and its relationship with other analytical channel models.

\subsection{Channel Model} \label{sec:Channel Model}

We consider a point-to-point mid-band XL-MIMO system, where the transmitter (Tx) and receiver (Rx) are equipped with $N_\rmT$ and $N_\rmR$ elements, respectively. 
As illustrated in Fig. \ref{fig:scenario}, we assume $L_{\rmR}$ and $L_\rmT$ clusters are present at the Rx and Tx sides, respectively.
Furthermore, each cluster at the Tx and Rx sides contains $Q_{\rmT,\ell_{\rmT}}$ and $Q_{\rmR,\ell_\rmR}$ scatterers, respectively.
For simplicity, we denote that $Q_{\rmT}=Q_{\rmT,\ell_\rmT}$ for $\ell_{\rmT}=1, \ldots,L_\rmT$ and $Q_{\rmR}=Q_{\rmR,\ell_\rmR}$ for $\ell_\rmR=1,\ldots,L_\rmR$.

In this paper, we assume a frequency-flat fading and time-invariant condition, and the channel derived from universal propagation mechanism is denoted as
\begin{equation}
    \bH = \sqrt{\mathsf{PL}} \sum_{\ell_\rmR=1}^{L_\rmR} \sum_{\ell_\rmT=1}^{L_\rmT}  \bH_{\ell_\rmR,\ell_\rmT},
    %\alpha_{\ell_\rmR,\ell_\rmT}
    \label{eq:Hgeneral}
\end{equation}
where $\sqrt{\mathsf{PL}}$ represents the path loss determined by the distance from the Tx to the Rx, $\bH_{\ell_\rmR,\ell_\rmT} \in \bbC ^{N_\rmR \times N_\rmT}$ is the channel from the Tx to the Rx, through the $\ell_\rmT$-th and $\ell_\rmR$-th clusters.
When a link does not exist between the $\ell_\rmR$-th cluster and the $\ell_\rmT$-th cluster, $\bH_{\ell_\rmR,\ell_\rmT} = \mathbf{0}$ is satisfied.
Note that in this paper, we mainly focus on the fast-fading characteristics of the mid-band XL-MIMO channel, and the long-term fading is included in the path loss term.

Clusters at the Tx side and the Rx side are considered as the first and last hops, respectively, and the propagation between the first and last interaction is not specified \cite{38901,GBSM1}.
Specifically, the $(n_\rmR,n_\rmT)$-th entry in $\bH_{\ell_\rmR,\ell_\rmT}$, denoted as $h_{n_\rmR,n_\rmT,\ell_\rmR,\ell_\rmT}$, can be modeled through electromagnetic propagation as \cite{25996} \footnote{In this paper, the geometric topology based near-field LoS component (like \cite{NFLoS,Fourier}) is not considered.
Without loss of generality, the LoS and NLoS components can be distinguished based on the link power.}
\begin{equation}%\small
	\begin{aligned}
		h_{n_\rmR,n_\rmT,\ell_\rmR,\ell_\rmT} \! =\! \sqrt{\frac{|\alpha_{\ell_\rmR,\ell_\rmT}|^2 \sigma_{\mathsf{SF}}}{Q_\rmR Q_\rmT}} \sum_{q_\rmR=1}^{Q_\rmR}\sum_{q_\rmT=1}^{Q_\rmT}e^{\jmath \phi_{\ell_\rmR,\ell_\rmT,q_\rmR,q_\rmT}}&\\
		\times \  g_{\ell_{\rmR},q_{\rmR}} \sqrt{G_{\rmR}(\Upsilon_{\ell_\rmR,q_\rmR})} \frac{\exp(-\jmath \kappa D_{n_\rmR}(\Upsilon_{\ell_\rmR,q_\rmR}))}{c ^{(n_\rmR)} _{\ell_{\rmR},q_{\rmR}}}&\\
		\times \  g_{\ell_{\rmT},q_{\rmT}} \sqrt{G_{\rmT}(\Upsilon_{\ell_\rmT,q_\rmT})} \frac{\exp(\jmath \kappa D_{n_\rmT}(\Upsilon_{\ell_\rmT,q_\rmT}))}{c ^{(n_\rmT)} _{\ell_{\rmT}, q_{\rmT}}},&
	\end{aligned}
 \label{eq:GBSM}
\end{equation}
where $|\alpha_{\ell_\rmR,\ell_\rmT}\vert ^2$ represents the power allocation of the link from the $\ell_{\rmT}$-th cluster to the $\ell_\rmR$-th cluster, satisfying $\sum _{\ell_\rmR=1} ^{L_\rmR} \sum _{\ell_\rmT = 1} ^{L_\rmT}|\alpha_{\ell_\rmR,\ell_\rmT}|^2=1$, $\sigma_{\mathsf{SF}}$ is the lognormal shadow fading (SF),   
$\phi_{\ell_\rmR,\ell_\rmT,q_{\rmR},q_{\rmT}}$ is the random phase shift introduced by the propagation and can be modeled as a uniform distribution $\mathcal{U}(-\pi,\pi)$, and $\kappa$ is the wave number.
For ease of introducing the remaining parameters in \eqref{eq:GBSM}, we use the subscript $\{\rmT,\rmR \}$ as ``${\bullet}$'' to indicate that the description applies to both the Tx and Rx sides. As a result, the parameters are as follows:
\begin{itemize}
    \item $g_{\ell_\bullet,q_\bullet}$ is the random complex gain satisfying $\sum _{q _\bullet} ^{Q _{\bullet}} \bbE \{|g_{\ell_\bullet,q_\bullet} \vert ^2\} / Q _{\bullet} =1$ as $Q _{\bullet} \rightarrow \infty$, which can be modeled as $g_{\ell_\bullet,q_\bullet} \sim \mathcal{CN}(0,1)$. 

    \item $G_{\bullet}(\Upsilon_{\ell_\bullet,q_\bullet})$ represents the antenna gain of each array element, which depends on the path incident to or reflected by the $q_\bullet$-th scatterer in the $\ell_\bullet$-th cluster located at $\Upsilon_{\ell_\bullet,q_\bullet}$. When uniform linear arrays (ULAs) are employed, $\Upsilon_{\ell_\bullet,q_\bullet}$ can be determined by $\{d_{\ell_\bullet,q_\bullet}, \theta_{\ell_\bullet,q_\bullet}\}$.\footnote{Note that we assume the deployment of ULAs for simplicity of analytical expression, and the following analysis also applies to other antenna topologies such as uniform planar antenna arrays (UPAs) \cite{NFUPA}.} 

    \item $D_{n_{\bullet}}(\Upsilon_{\ell_{\bullet},q_{\bullet}})$ is the distance between the scatterer with parameter $\Upsilon_{\ell_{\bullet},q_{\bullet}}$ to the $n_{\bullet}$-th array element, given by 
\begin{multline}
        D_{n_\bullet}(\Upsilon_{\ell_\bullet,q_\bullet}) \! =\! \sqrt{d^2_{\ell_{\bullet},q_\bullet} \! - \! 2\eta_{\bullet,n_\bullet} d_{\ell_{\bullet},q_\bullet}  \sin \theta_{\ell_{\bullet},q_\bullet} \! + \! \eta^2_{\bullet,n_\bullet} } \\
        \approx \! d_{\ell_{\bullet},q_\bullet} \!-\! \eta_{\bullet,n_\bullet} \sin \theta_{\ell_{\bullet},q_\bullet} \!+\! \frac{\eta ^2 _{\bullet,n_\bullet} }{2d_{\ell _\bullet,q _\bullet}} {\left(1 \! - \! \sin ^2 \theta _{\ell _\bullet,q _\bullet} \right)}, 
        \label{eq:dist}     
\end{multline}
    where the approximation employs Taylor's series expansion for $\sqrt{1+x}\approx 1 +\frac{1}{2}x-\frac{1}{8}x^2$, commonly referred to as the Fresnel approximation \cite{Fresnel}. Here, $\eta_{\bullet,n_\bullet}$ represents the distance between the $n_\bullet$-th antenna and the center of the array at either the Tx or the Rx. 

    \item {$c ^{(n _\bullet)} _{\ell _\bullet,q _\bullet}$ captures the non-stationary characteristic along the extra large-scale array, and we set $c ^{(n _\bullet)} _{\ell _\bullet,q _\bullet} \propto D _{n_\bullet} (\Upsilon _{\ell _\bullet,q _\bullet}) / d _{\ell _\bullet,q _\bullet}$.
    Note that $c ^{(n _\bullet)} _{\ell _\bullet,q _\bullet}$ is not equivalent to the large-scale fading, i.e., $\sqrt{\mathsf{PL}}$, where the former captures the power fluctuation along array elements, and the latter is decided by the distance between the Tx and the Rx.} 
\end{itemize}

\subsection{Proposed Analytical Channel Model}

Although comprehensive, the complexity of the channel described by \eqref{eq:GBSM} presents significant challenges for direct theoretical analysis. To mitigate this, we develop an analytical channel model by further abstracting certain features outlined in the previous subsection. Because we specifically concentrate on the spatial correlation associated with small-scale fading, as defined in \eqref{eq:GBSM}, we deliberately exclude considerations of shadow fading to simplify our analysis. Moreover, we assume $G_{\bullet}(\Upsilon_{\ell_{\bullet},q_{\bullet}})$ for all $\ell_{\bullet}$ and $q_{\bullet}$ to be unity when using omnidirectional antenna elements at both Tx and Rx arrays. 

To further simplify the model, we introduce a key assumption: given the probable distance between receiver clusters and transmitter clusters, we can statistically model the propagation link between any cluster at the Tx and any cluster at the Rx. This simplification allows us to represent the channel from the cluster $\ell_\rmT$ to the cluster $\ell_\rmR$, denoted as $\bH _{\ell_\rmR,\ell_\rmT}$ in \eqref{eq:Hgeneral}, as the product of two vectors: $\bH_{\ell_\rmR,\ell_\rmT} = \bh_{\rmR,\ell_\rmR} \bh_{\rmT,\ell_\rmT} ^ {\ctrans}$. Here, $\bh_{\rmT,\ell_\rmT} \in \bbC ^{N_\rmR \times 1}$ represents the rays traveling from the Tx to cluster $\ell_\rmT$, and  $\bh_{\rmR,\ell_\rmR} \in \bbC ^{N_\rmR \times 1}$ represents the rays traveling from cluster $\ell_\rmR$ to the Rx. The proposed channel model is thus formulated as
\begin{equation}
	\bH = \sqrt{\mathsf{PL}} \sum_{\ell_{\rmR}=1}^{L_{\rmR}} \sum_{\ell_{\rmT}=1}^{L_{\rmT}} \alpha_{\ell_{\rmR},\ell_{\rmT}} \bh_{\rmR,\ell_{\rmR}} \bh^{\ctrans}_{\rmT,\ell_{\rmT}}
	 = \sqrt{\mathsf{PL}} \bH_{\rmR} \bA \bH_{\rmT}^{\ctrans},\label{eq:channel}
\end{equation}
where $\bA\in \bbC^{L_\rmR \times L_\rmT}$ denotes the power allocation among different link coupling relationships, with $[\bA]_{\ell_{\rmR},\ell_\rmT}=\alpha_{\ell_\rmR,\ell_\rmT}$.  Matrices $\bH_\rmR \in \bbC^{N_{\rmR}\times L_\rmR}$ and $\bH_\rmT \in \bbC^{N_\rmT \times L_\rmT}$ are represented as $\bH_\rmR=[\bh_{\rmR,1},\ldots,\bh_{\rmR,L_\rmR}]$ and $\bH_\rmT = [\bh_{\rmT,1},\ldots, \bh_{\rmT,L_\rmT}]$, respectively.
Notably, \eqref{eq:channel} still preserves the essential signal propagation mechanisms between clusters near both the Tx and Rx, ensuring alignment with the critical characteristics highlighted in \ref{sec:XLMIMOfeature} and maintaining focus on these dynamics at the cluster level. 
Additionally, \eqref{eq:channel} is also applicable to scenarios where multiple rays exist between a cluster at the Tx and a cluster at the Rx side. In such cases, $\bH _\rmR$, $\bH _\rmT$ and $\bA$ can be derived by stacking the contributions of multiple rays at the Rx and Tx respectively.

To account for spatial correlation properties, $\bh_{\rmR,\ell_{\rmR}}$ and $\bh_{\rmT,\ell_\rmT}$ are described statistically as $\cov_{\rmR, \ell_{\rmR}} = \bbE \{\bh_{\rmR,\ell_{\rmR}} \bh ^{\ctrans} _{\rmR,\ell_{\rmR}} \}$, $\cov_{\rmT, \ell_{\rmT}} = \bbE \{\bh_{\rmT,\ell_{\rmT}} \bh ^{\ctrans} _{\rmT,\ell_{\rmT}} \}$, respectively, where $\cov_{\rmR, \ell_{\rmR}} \in \bbC ^{N_ \rmR \times N _\rmR}$ and $\cov_{\rmT, \ell_{\rmT}} \in \bbC ^{N _\rmT \times N _\rmT}$ are the spatial correlation matrices for cluster $\ell_{\rmR}$ and cluster $\ell_{\rmT}$, respectively. 
{Furthermore, $\bh_{\rmR,\ell_\rmR}$ and $\bh_{\rmT,\ell_\rmT}$ are expressed via the Karhunen-Lo\`{e}ve representation \cite{KL} as follows:
\begin{subequations}
\label{eq:hRhT}
\begin{align}
\bh_{\rmR,\ell_\rmR} &= \cov_{\rmR,\ell_\rmR}^{1/2} \tilde{\bg}_{\rmR,\ell_{\rmR}} = \bU _{\rmR,\ell_\rmR} \bLambda _{\rmR,\ell_\rmR} ^{1/2} {\bg}_{\rmR,\ell_{\rmR}},\\
\bh_{\rmT,\ell_\rmT} &= \cov_{\rmT,\ell_\rmT}^{1/2} \tilde{\bg}_{\rmT,\ell_{\rmT}} = \bU _{\rmT,\ell_\rmT} \bLambda _{\rmT,\ell_\rmT} ^{1/2} {\bg}_{\rmT,\ell_{\rmT}},
\end{align}
\end{subequations}
where $\bU _{\rmR,\ell_\rmR}$ and $\bU _{\rmT,\ell_\rmT}$ contain the eigenvectors of $\cov_{\rmR,\ell_\rmR}$ and $\cov_{\rmT,\ell_\rmT}$, respectively, $\bLambda _{\rmR,\ell_\rmR}$ and $\bLambda _{\rmT,\ell_\rmT}$ are diagonal matrices containing eigenvalues of $\cov_{\rmR,\ell_\rmR}$ and $\cov_{\rmT,\ell_\rmT}$, respectively, $\tilde{\bg} _{\rmR,\ell_{\rmR}},\ \bg_{\rmR, \ell_{\rmR}} \in \bbC^{N_\rmR\times 1}$ and $\tilde{\bg} _{\rmT,\ell_\rmT},\ \bg_{\rmT,\ell_\rmT} \in \bbC^{N_\rmT \times 1}$ are random vectors with entries satisfying $\mathcal{CN}(0,1)$.}
To model the spatial correlation matrices $\cov_{\bullet,\ell _{\bullet}}$, we first derive the following proposition.
\begin{proposition}
\label{prop:kronecker}
    The joint spatial correlation matrix from cluster $\ell_\rmT$ to cluster $\ell_\rmR$, denoted as $\cov _{\ell_\rmR, \ell_\rmT}$, satisfies
\begin{equation}
\setlength\abovedisplayskip{3pt}
\setlength\belowdisplayskip{3pt}
    \cov_{\ell_\rmR,\ell_\rmT} = \cov_{\rmR,\ell_\rmR} \otimes \cov_{\rmT,\ell_\rmT}.
    \label{eq:prop1}
\end{equation}
\end{proposition}
\begin{IEEEproof}
    Refer to Appendix \ref{proof:kronecker}.
\end{IEEEproof}
{\it Proposition} \ref{prop:kronecker} allows us to model the statistical spatial correlation characteristics at the Tx and Rx separately. Note that unlike the widely used Kronecker channel model \cite{Kronecker}, the Kronecker property in the proposed channel model is satisfied for propagation links between a Tx cluster and a Rx cluster rather than the entire channel. This distinction endows our model with greater applicability and flexibility.

Based on \eqref{eq:xiX} in Appendix \ref{proof:kronecker}, and considering the number of scatterers in a cluster tends towards infinity, i.e., $Q_{\bullet} \rightarrow \infty$, the spatial correlation coefficient between the $n_{\bullet_1}$-th array element and the $n_{\bullet_2}$-th array element, denoted by $\xi_{n_{\bullet_2}}^{(n_{\bullet_1})}$, can be further modeled statistically based on the distribution of parameters as follows: 
\begin{multline}
    \xi_{n_{\bullet_2}}^{(n_{\bullet_1})} =  \int  \frac{\exp \left(\jmath \kappa D_{n_{\bullet_2}}(\Upsilon_{\ell_\bullet,q_\bullet}) \right)}{ c ^{(n_{\bullet_2})} _{\ell_{\bullet}, q_{\bullet}} } \\
    \times 
    \frac{\exp \left( -\jmath \kappa D_{n_{\bullet_1}}(\Upsilon_{\ell_\bullet,q_\bullet})\right)}{c ^{(n_{\bullet_1})} _{\ell_{\bullet}, q_{\bullet}}} 
    f(\Upsilon_{\ell_\bullet,q_\bullet}) \, \rmd \Upsilon_{\ell_\bullet,q_\bullet}, \label{eq:NFCor}
\end{multline}
% _{\Upsilon _{\ell_\bullet,q_\bullet}}
due to
$\bbE\{|g_{\ell_\bullet,q_\bullet}|^2\}/Q_\bullet = f(\Upsilon _{\ell_\bullet,q_\bullet}) \, \rmd \Upsilon_{\ell_\bullet,q_\bullet}$ \cite{onering}, where $f(\Upsilon_{\ell_{\bullet}, q_{\bullet}})$ represents a distribution of $\Upsilon_{\ell_{\bullet}, q_{\bullet}}$, i.e., the probability distribution function (PDF) of $\Upsilon_{\ell_{\bullet}, q_{\bullet}}$. 
Generally, $d _{\ell _\bullet,q_\bullet}$ and $\theta _{\ell _\bullet,q_\bullet}$ can be modeled independently, and $f(\Upsilon _{\ell_{\bullet}, q_{\bullet}})$ can be written as $f(d _{\ell _\bullet,q_\bullet}) f(\theta _{\ell _\bullet,q_\bullet})$ \cite{dist1,dist2}.
However, the variation of $d _{\ell _\bullet,q _\bullet}$ can be considered negligible due to the approximate constancy of scatterer ranges within a cluster. Therefore, we simplify $f(\Upsilon_{\ell_{\bullet}, q_{\bullet}})$ to $f(\theta_{\ell_{\bullet}, q_{\bullet}})$ by assuming $d _{\ell _\bullet,q _\bullet}$ as a constant.
Furthermore, the Von Mises distribution (VMD) can be adopted for modeling $f(\theta_{\ell_{\bullet}, q_{\bullet}})$\footnote{Note that the VMD is a widely used distribution to describe the angular spread, which can describe the degree of angular spread flexibly through changing $\angwid$, other distribution models can also be adopted. When the UPAs are deployed, the Von Mises Fisher distribution can be adopted.} with subscripts omitted
\begin{equation}
    f(\theta) = \frac{\exp\left(\angwid ^{-1} \cos(\theta - \mu_{\theta})\right)}{2\pi I_0(\angwid ^{-1})},
\end{equation}
where $\mu_{\theta}$ is the mean angle, $\angwid > 0$ controls the width of the angular distribution with $\angwid \rightarrow 0$ giving a ray at the single angle $\mu_{\theta}$ and $\angwid\rightarrow \infty$ giving a uniform spread of angles, and $I_0(\cdot)$ is the zeroth-order modified Bessel function of the first kind. 
Note that in the proposed model, the near-field properties are incorporated with the spatial correlation modeling, and the influence of near-field characteristics is further specified in Section \ref{sec:SE}.

\begin{remark}
One may notice that the proposed channel model is similar to the \emph{multi-keyhole} channel model \cite{multikeyhole1}. However, in this paper, the $\bA$ is not restricted to a diagonal matrix, to describe a more complex scattering environment. Moreover, the wide applicability of the multi-keyhole model is also inherited and the proposed channel model can embrace a variety of typical analytical channel models, as well as encompass the far-field and near-field situations alike, which will be specified in Section \ref{subsec:discuss}.
\end{remark}

\subsection{Compatibility Analysis}
\label{subsec:discuss}
The proposed channel model demonstrates high flexibility and compatibility, enabling the incorporation of a broad frequency range and various application scenarios for mid-band XL-MIMO systems, alongside other conventional analytical channel models.
We further elucidate the model's compatibility through the following three aspects.

\subsubsection{Compatibility with Various Scenarios}
The proposed channel model adeptly characterizes the propagation environment and cluster characteristics, as outlined below:
\begin{itemize}
\item Near-field propagation characteristics are statistically modeled at the cluster level, with added consideration for the clusters' central distances compared to the far-field model.
\item The propagation environment's features (e.g., sparse or dense scattering) are depicted through the power coupling matrix $\bA$.
\item Beyond the aforementioned channel characteristics, other channel characteristics such as path loss and angular spread are also adjustable through parameter settings to accommodate different frequency bands and application scenarios. Furthermore, various measurement results can be seamlessly integrated for channel parameterization.
\end{itemize}

\subsubsection{Compatibility with Other Analytical Models}

Next, we examine the compatibility of our proposed analytical model with other analytical channel models. Intuitively, as the angular spread becomes exceedingly small, that is, $\angwid \rightarrow 0$, our channel model simplifies to a finite-dimensional channel model. This simplification enables the characterization of the channel from the perspective of propagation rays.
In terms of statistical characteristics and akin to the multi-keyhole model, our model aligns with the double-scattering channel model \cite{double} under conditions where $\cov_{\rmR,\ell_\rmR} = \cov_{\rmR}, \forall \ell_\rmR$ and $\cov_{\rmT,\ell_\rmT} = \cov _{\rmT}, \forall \ell_\rmT$ are met, and when $\bA$ is a Hermitian matrix with uniform diagonal elements. This compatibility extends to the correlated Rayleigh model when $\bA = \bI$ and $L_{\rmR}$ and $L_\rmT$ approach infinity.
Furthermore, when regarding the eigenmode channel, the eigenmodes at the Tx and Rx are identified as follows:
\begin{proposition}\label{prop:eigenmode}
    The eigenmodes of the Tx and Rx, denoted by $\bU_{\rmT}$ and $\bU_{\rmR}$, respectively, are derived through the eigenvalue decomposition of the spatial correlation matrices at both link ends, i.e., $\bbE\{\bH ^{\ctrans} \bH\}$ and $\bbE\{\bH \bH ^{\ctrans}\}$. The decomposition is given by:
    \begin{subequations}
    \label{eq:eigenmode}
    \begin{align}
        \bU_{\rmT}\bLambda_{\rmT}\bU_{\rmT}^{\ctrans} &  \approx \sum_{\ell_\rmT=1} ^{L_\rmT} \sum_{\ell_\rmR=1}^{L_\rmR} |\alpha_{\ell_\rmR,\ell_\rmT}|^2 \Tr(\cov_{\rmR,\ell_\rmR})  \cov_{\rmT,\ell_\rmT} ,\\
        \bU_{\rmR}\bLambda_{\rmR}\bU_{\rmR}^{\ctrans} &  \approx \sum_{\ell_\rmT=1} ^{L_\rmT} \sum_{\ell_\rmR=1}^{L_\rmR} |\alpha_{\ell_\rmR,\ell_\rmT}|^2 \Tr(\cov_{\rmT,\ell_\rmT}) \cov_{\rmR,\ell_\rmR}.
    \end{align}
    \end{subequations}
\end{proposition}
\begin{IEEEproof}
    Refer to Appendix \ref{proof:eigenmode}.
\end{IEEEproof}
Based on the analysis above, it is evident that the proposed channel model can be adapted to other analytical channel models, which also reflects the versatility of the proposed model in characterizing various channel propagation environments.

\subsubsection{Compatibility with Far-Field and Near-Field}
Next, we delve into the model's compatibility with far-field conditions. When the number of antennas is small, the effects of spherical wavefront become negligible, allowing the distance in \eqref{eq:dist} to be approximated using Taylor series expansion as
\begin{equation}
    D_{n_\bullet}(\Upsilon _{\ell_\bullet,q_{\bullet}}) \approx d_{\ell_\bullet, q_{\bullet}} - \eta_{\bullet, n_\bullet} \sin\theta _{\ell_\bullet,q _\bullet}.
\end{equation}
Hence, the model can be simplified to a far-field scenario, particularly apt for situations where the UE is equipped with a limited number of antennas. Additionally, the spatial correlation matrix in the far field can be expressed as
\begin{equation}
\begin{aligned}
    [&\cov_{\bullet,\ell_\bullet} ]_{n_{\bullet_1},n_{\bullet_2}}
     = \\
     &{ I_0 \left(\sqrt{\angwid^{-2} \! + \! b^2_{n_{\bullet_1}n_{\bullet_2}} \! + \! 2b _{n_{\bullet1}n_{\bullet2}}\angwid ^{-1} \sin \mu _{\theta_{\ell_\bullet}}} \right)}/{I_0(\angwid ^{-1})},
\end{aligned}
    \label{eq:cov_FF}
\end{equation}
where $b_{n_{\bullet_1} n_{\bullet_2}}=\jmath \kappa(\eta_{\bullet, n_{\bullet_1}} - \eta_{\bullet, n_{\bullet_2}})$.
It can be obtained from \eqref{eq:cov_FF} that $\cov_{\bullet, \ell_ \bullet} $ is a Toeplitz and Herimitian matrix under far-field conditions.
Therefore, the far-field spatial correlation matrix can asymptotically be diagonalized using a DFT matrix \cite{ToepDFT,FFGao}, thereby setting the eigenmode bases as DFT matrices.

\section{Performance Analysis for Mid-Band XL-MIMO Systems}
To address the problem {\bf P2} mentioned in the Introduction, we investigate the performance of mid-band XL-MIMO systems based on the proposed channel model. Our primary focus is on the ergodic SE and the OP-MRC, which reflect two fundamental requirements: transmission throughput and reliability, respectively. Specifically, for convenience of analysis, two typical scenarios are selected. Closed-form approximations and performance bounds are derived, while the characteristics and impacts of mid-band XL-MIMO systems are analyzed based on these derived expressions.

\subsection{Metrics of Mid-Band XL-MIMO Systems}
Throughout the historical review, advanced transmission schemes have been proposed to explore channel characteristics \cite{SMBF}. To explore the performance of mid-band XL-MIMO systems, spatial multiplexing (SM) and beamforming (BF) are assumed, and the corresponding ergodic SE and OP-MRC are selected as two effective performance indicators.

\subsubsection{Ergodic SE under SM} 
Assuming the transmit signal is $\mathbf{x}\in \bbC^{N_\rmT\times 1}$, the signal received at the Rx is given by 
\begin{equation}
    \by = \bH \bx + \bn,
\end{equation}
where $\bn \in \bbC^{N_\rmR\times 1}$ is the additive white Gaussian noise (AWGN) with zero means and covariance $\iota_n^2 \bI$, and we assume $\iota_n^2 = 1$ for simplicity.
Considering the multiplexing transmission is adopted and an equal power allocation strategy is employed, the covariance of the transmit signal is $\bQ = \bbE\left\{ \bx \bx^{\ctrans} \right\} = \frac{P}{N_\rmT} \bI_{N_\rmT}$,
where $P=\bbE\{\Tr(\bx \bx^{\ctrans}) \}$ is the total transmit power. The ergodic SE is denoted as
\begin{align}
    C & = \bbE_{\bH} {\left\{\log_2\det\left(\bI + \gamma \bH \bH^{\ctrans}\right) \right\}}, \notag \\
    & = \bbE_{\blambda(\bH \bH ^{\ctrans})} {\left\{\sum_{i=1}^{\rank(\bH)}\log_2\left(1+\gamma\lambda_i(\bH \bH ^{\ctrans}) \right) \right\}},
    \label{eq:capacity}
\end{align}
where $\gamma = P/(N_\rmT\iota_n^2)$, and $\lambda_i (\bH \bH ^{\ctrans})$ is the $i$-th eigenvalue of $\bH \bH^{\ctrans}$.

\subsubsection{OP under BF}

When the BF transmission strategy is adopted, we focus on the OP, which reflects the reliability of transmission. Considering a MIMO system with MRC, the estimate of the single-stream transmit signal $x$ is given by $\hat{x} = \sqrt{\bar{\gamma}} \bw^{\ctrans} \bH^{\ctrans} ( \bH \bw x + \bn)$, where $\bw \in \bbC^{N_{\rmT} \times 1}$ is the beamforming vector satisfying $\bw^{\ctrans} \bw=1$, and $\bar{\gamma}$ denotes the ratio of transmit power to the noise variance $\iota_n^2$. The receive SNR is then ${\gamma} _{\rmR} = \bar{\gamma} \bw^{\ctrans} \bH^{\ctrans} \bH \bw$.
It is well known that the BF vector $\bw_{\opt}$, which maximizes the receive SNR, is the eigenvector corresponding to the maximum eigenvalue of $\bH^{\ctrans} \bH$.
Assuming perfect CSI and the BF vector $\bw_{\opt}$ are available at both the Rx and the Tx, the receive SNR is further given by
\begin{equation}
    \gamma_{\rmR}=\bar{\gamma} \bw_{\opt}^{\ctrans} \bH^{\ctrans} \bH \bw _{\opt} = \bar{\gamma} \lambda_{\max} (\bH^{\ctrans} \bH).
\end{equation}
The OP is defined as the probability that the receive SNR, $\gamma_{\rmR}$, drops below an acceptable SNR threshold $\gamma_{\threshold}$, i.e.,
\begin{equation}
    P_{\mathsf{out}} = \Prob{\left(\gamma_\rmR \leq \gamma_{\threshold} \right)}=\Prob {\left(\lambda_{\max}(\bH ^{\ctrans} \bH) \leq \frac{\gamma_{\threshold}}{\bar{\gamma}} \right)}.
\end{equation}

\subsection{Typical Scenarios}
Given the broad applicability and compatibility of the proposed channel model, directly analyzing the aforementioned metrics through the general expression in \eqref{eq:channel} proves challenging. Therefore, we select two typical scenarios aimed at representing the characteristics of the mid-band channel under both the high and low parts of the mid-band spectrum, respectively, for specific illustrations. General cases are analyzed through simulations in the following section. Without loss of generality, the path loss term $\sqrt{\mathsf{PL}}$ is omitted in this section.

\subsubsection{Specular Components Dominant Scattering (SS) Scenario}
Considering a scenario dominated by specular components, the angular spread is assumed to be extremely narrow. Thus, $\angwid$ approaches 0, and the rank of $\cov_{\bullet,\ell_\bullet}$ satisfies $\rank(\cov_{\bullet,\ell_\bullet})\approx 1$. Consequently, based on \eqref{eq:hRhT}, $\bh_{\bullet, \ell _\bullet}$ can be further approximated as
\begin{equation}
    \bh_{\bullet,\ell_\bullet} = \bU_{\bullet,\ell_\bullet} \bLambda _{\bullet,\ell_\bullet} ^{\half} \bg_{\bullet,\ell_\bullet} \approx \sqrt{\chi_{\bullet,\ell_\bullet}} g_{\bullet,\ell_\bullet} \bu_{\bullet,\ell_\bullet},
    \label{eq:h_SS_app}
\end{equation}
where $\chi_{\bullet,\ell_\bullet}$ is the maximum eigenvalue of $\cov_{\bullet,\ell_\bullet}$, $\bu_{\bullet,\ell_\bullet}$ is the eigenvector corresponding to $\chi_{\bullet,\ell_\bullet}$, and $g_{\bullet,\ell_\bullet}$ is an complex Gaussian variable satisfying $\mathcal{CN}(0,1)$.
Furthermore, in communication with higher frequency bands of the mid-band spectrum, large-scale fading becomes severe, and several clusters are distinguishable at both the Tx side and the Rx side, thus resulting in a sparse propagation environment, i.e., a sparse $\bA$ under SS scenarios. Under these circumstances, the equivalent channel expression is derived in the following proposition.
 
\begin{proposition}\label{prop:sparsity}
    In the SS scenario, the proposed channel model is equivalent to the following expression:
    \begin{equation}
        \bH_{\sparse} = \bPi_\rmR \boldsymbol{\Lambda}_\rmR \bA \bLambda^{\ctrans}_{\rmT} \bPi^{\ctrans}_\rmT = \bPi_\rmR \tilde{\bA} \bPi^{\ctrans}_{\rmT},
        \label{eq:sparsescatter}
    \end{equation}
    where $\bPi_\rmR = [\bu_{\rmR,1},\ldots,\bu_{\rmR,L_\rmR}]$, $\bPi_\rmT = [\bu_{\rmT,1},\ldots,\bu_{\rmT,L_\rmT}]$, $\bLambda_\rmR=\diag(\sqrt{\chi_{\rmR,1}} g_{\rmR,1},\ldots,\sqrt{\chi_{\rmR,L_\rmR}} g_{\rmR,L_\rmR})$, $\bLambda_\rmT = \diag(\sqrt{\chi_{\rmT,1}} g_{\rmT,1},\ldots,\sqrt{\chi_{\rmT,L_\rmT}} g_{\rmT,L_\rmT})$, and $\tilde{\bA} = \bLambda_\rmR \bA \bLambda _{\rmT} ^{\ctrans}$.
\end{proposition}
\begin{IEEEproof}
    Based on \eqref{eq:h_SS_app}, $\bH _{\bullet}$, in \eqref{eq:channel} can be further denoted as 
    \begin{equation}
    \begin{aligned}
        \bH _{\bullet} &\!= \! [\sqrt{\chi_{\bullet,1}} g_{\bullet,1} \bu_{\bullet,1},\dots,\sqrt{\chi_{\rmR,L_\bullet}} g_{\rmR,L_\bullet} \bu_{\bullet,L_\bullet}]\\
        &\! =\! [\bu_{\bullet,1},\dots,\bu_{\bullet,L_\bullet}] \diag(\chi ^{\half}_{\bullet,1} g_{\bullet,1}, \ldots, {\chi ^{\half} _{\bullet,L_\bullet}} g_{\bullet,L_\bullet}).
    \end{aligned}
    \end{equation}
    Therefore, $\bPi_\rmR$, $\bPi_\rmT$, $\bLambda_\rmR$ and $\bLambda_\rmT$ can be defined.
\end{IEEEproof}

Note that the expression in \eqref{eq:sparsescatter} is equivalent to the finite-dimensional channel model in a near-field form \cite{finite,NFfinite}, denoted as
\begin{equation}
\bH_{\sparse} = \bB_{\rmR} \breve{\bA} \bB^{\ctrans}_{\rmT},
\end{equation}
where $\bB_{\rmR}=[\bb_{\rmR}(\Upsilon_{1}),\ldots,\bb_{\rmR}(\Upsilon_{L_{\rmR}})]$, and $\bB_{\rmT}=[\bb_{\rmT}(\Upsilon_{1}),\dots,\bb_{\rmT}(\Upsilon_{L_{\rmT}})]$, $\bb_{\bullet}(\Upsilon_{\ell_\bullet})$ is the near-field steering vector, satisfying $[\bb_{\bullet}(\Upsilon_{\ell_\bullet})]_{n_\bullet} = \exp (-\jmath \kappa D_{n_\bullet}(\Upsilon_{\ell_\bullet})) / c ^{(n _\bullet)} _{\ell _\bullet}$,  
$\Upsilon _{\ell _\bullet} = \{d _{\ell _\bullet}, \theta _{\ell _\bullet}\}$, with $d _{\ell _\bullet}$ and $\theta _{\ell _\bullet}$ representing the distance and mean angle of the center of cluster $\ell _\bullet$, and $[\breve{\bA}] _{\ell_\rmR, \ell_\rmT} = \alpha _{\ell_\rmR, \ell_\rmT} {g}_{\rmR,\ell_\rmR} g^{*}_{\rmT,\ell_\rmT}$.

\subsubsection{Dense Components Dominant Scattering (DS) Scenario}
We then consider the scenario dominated by dense multipath components where the angular spread is significant. Meanwhile, assuming the system operates within the lower mid-band, and both the Tx and Rx are surrounded by a complex scattering environment, $\bA$ is not a sparse matrix, reflecting complex scattering relationships between different clusters. Under this circumstance, the approximate channel representation is derived in the following proposition.
\begin{proposition}\label{prop:richscatter}
In a dense components dominant scattering scenario, the channel is approximated by
\begin{equation}
\bH _{\RS} \approx \bar{\cov}^{\half}_{\rmR} \bH_{w,\rmR} \bA \bH_{w,\rmT} ^{\ctrans} \bar{\cov}^{\frac{\ctrans}{2}}_{\rmT},
\label{eq:doublescatter}
\end{equation}
where $\bH_{w,\rmR}\in \bbC^{N_{\rmR}\times L_\rmR}$ and $\bH_{w,\rmT}\in \bbC^{N_\rmT \times L_\rmT}$ are random Gaussian matrices with i.i.d. $\CN(0,1)$ elements. $\bar{\cov}_{\rmR}$ and $\bar{\cov}_{\rmT}$ satisfy
\begin{subequations}
\label{eq:barCov}
\begin{align}
    \bar{\cov}_{\rmR}&=\frac{1}{L_{\rmR}}(\cov_{\rmR,1} + \dots + \cov_{\rmR,L_{\rmR}}),\\
    \bar{\cov}_{\rmT}&=\frac{1}{L_{\rmT}}(\cov_{\rmT,1} + \dots + \cov_{\rmT,L_{\rmT}}).
\end{align}
\end{subequations}
\end{proposition}
\begin{IEEEproof}
 See Appendix \ref{proof:richscatter}.
\end{IEEEproof}

\begin{remark}
The physical interpretation of \textit{Proposition} \ref{prop:richscatter} suggests that although the $\angwid$ at the Tx and the Rx are relatively small, the incident signal from  $L_\rmR$ clusters and the transmit signal to $L_\rmT$ clusters can be considered as being distributed across various directions with comparable power, which can be regarded as the additivity of the power angular spectrum.
Note that the right side of \eqref{eq:doublescatter} bears similarity to the double-scattering channel model, even if spatial correlation matrices are modeled at the cluster level at both Tx and Rx separately. Furthermore, the pinhole effect \cite{keyhole} can also be represented when $\rank(\bA)=1$.
\end{remark}

Note that the aforementioned two typical scenarios may both appear in mid-band MIMO systems, while the former is also suitable for mmWave MIMO systems, and the latter is effective in characterizing MIMO systems operating in the Sub-6 GHz frequency band. 
Furthermore, the sparse scattering and rich scattering can also be depicted by the SS scenario and the DS scenario above.
To sum up, these two equivalent expressions provide a basis for subsequent performance comparisons of different systems.

\subsection{Ergodic SE Analysis}
\label{sec:SE}

\subsubsection{Ergodic SE in SS Scenario}
Under the SS scenario, we derive both a closed-form approximation and an upper bound based on the equivalent channel in \eqref{eq:sparsescatter}. For simplicity, we assume that no more than one non-zero element exists in each row and column of $\bA$, defining $L_{\sparse}=L_\rmR = L_\rmT =\rank(\bH_{\sparse})$.\footnote{Note that this analysis and results also apply to the case where $L_\rmR \neq L_\rmT$, because $\tilde{\bA}$ can be rearranged into a diagonal matrix based on the aforementioned assumption.}

\begin{theorem}
\label{theorem:capacitysparse}
For a mid-band XL-MIMO system with equal power allocation, the approximation of the ergodic SE under the SS scenario is given by
\begin{multline}
\label{eq:capacityapp_sparse}
    C_{\sparse} ^{\app} = \frac{1}{\ln 2} \\
     ~ \sum_{\ell=1}^{L_\sparse} G_{4,2} ^{1,1} {\left( \gamma |\alpha_\ell|^2 \chi_{\rmR,\ell}\chi_{\rmT,\ell} \lambda_{\rmR,\ell}\lambda_{\rmT,\ell} \Big| \begin{array}{cccc} 0&0&1&1 \\ 1&0&& \end{array}\right)},
\end{multline}
where $\ell = 1,\ldots,L_{\sparse}$, $\lambda_{\bullet,\ell}$ is the $\ell$-th eigenvalue of $\bPi_{\bullet}^{\ctrans} \bPi_{\bullet}$, and $G_{p,q}^{m,n} \left( \cdot | \cdot \right)$ is the Meijer-G function. 
\end{theorem}
\begin{IEEEproof}
    See Appendix \ref{proof:capacitysparseapp}.
\end{IEEEproof}
Despite the approximation's tightness, its complexity hinders a thorough exploration of the impact introduced by mid-band XL-MIMO systems. To provide more insightful guidelines and capture the performance scaling introduced by the novel characteristics, we offer the following upper bound.
\begin{proposition}
\label{prop:sparse}
    For a mid-band XL-MIMO system with equal power allocation, the ergodic SE under the SS scenario can be upper bounded by
    \begin{equation}
    \label{eq:capacitysparseub}
        C_{\mathsf{SS}}\leq C_{\mathsf{\sparse}}^{\ub} = \sum_{\ell=1}^{L_{\sparse}}\log_2 {\left(1 + \gamma |\alpha_\ell|^2 \chi_{\rmR,\ell}\chi_{\rmT,\ell} \lambda_{\rmR,\ell}\lambda_{\rmT,\ell} \right)}.
    \end{equation}
\end{proposition}
\begin{IEEEproof}
    Applying Jensen's inequality to the ergodic SE in \eqref{eq:capacitysparseapp1}, we obtain
    \begin{equation}
    \begin{aligned}
        C_{\sparse} & \leq  \sum_{\ell=1}^{L_{\sparse}}\log_2 {\left(1 \! + \! \gamma |\alpha_\ell|^2 \chi_{\rmR,\ell}\chi_{\rmT,\ell} \lambda_{\rmR,\ell}\lambda_{\rmT,\ell} \bbE\left\{ |g_{\rmR,\ell} g^{*}_{\rmT,\ell}|^2 \right\} \right)} \\
        & \overset{(a)}{=} \sum_{\ell=1}^{L_{\sparse}}\log_2 {\left(1 \! + \! \gamma |\alpha_\ell|^2 \chi_{\rmR,\ell}\chi_{\rmT,\ell} \lambda_{\rmR,\ell}\lambda_{\rmT,\ell} \right)} = C_{\sparse} ^{\ub},
    \end{aligned}
    \label{eq:C_SS_ub}
    \end{equation}
    where (a) leverages the property that for independent $|g_{\rmR,\ell}|^2$ and $|g_{\rmT,\ell}|^2$, following an exponential distribution, $\bbE \{ |g_{\rmR,\ell} g^{*}_{\rmT,\ell}|^2 \} = \bbE \{ |g_{\rmR,\ell}|^2  \} \bbE \{ |g_{\rmT,\ell}|^2 \}=1$. 
\end{IEEEproof}

\begin{figure}[!t]
    \centering
    \includegraphics[scale=0.55]{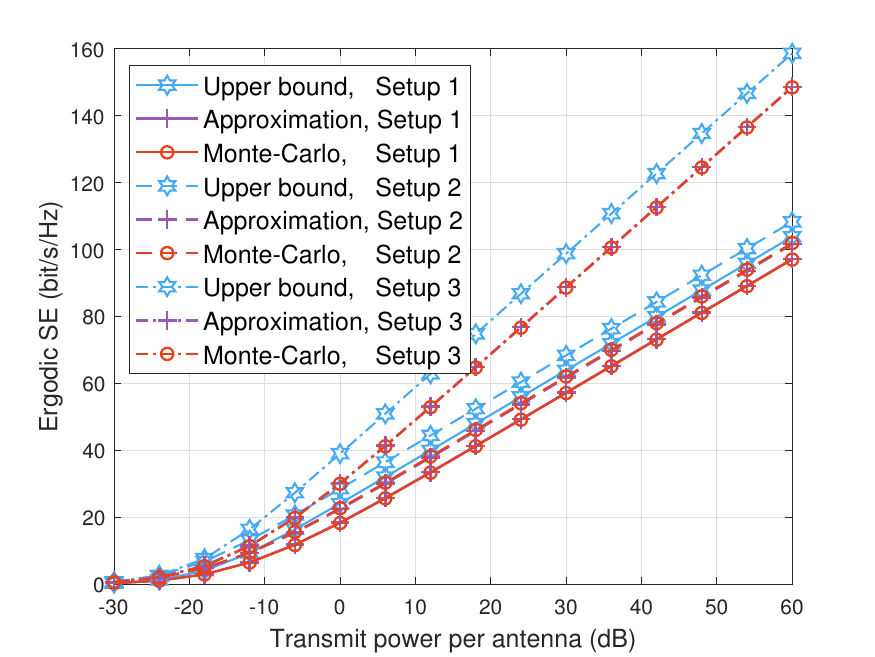} %0.45
    \caption{Ergodic SE of SS scenario against the transmit power per antenna. Setup 1: $N_\rmR = 256$, $N_\rmT=16$, $L_{\sparse}=4$; Setup 2: $N_\rmR = 512$, $N_\rmT=64$, $L_{\sparse}=4$; Setup 3: $N_\rmR = 512$, $N_\rmT=64$, $L_{\sparse}=6$.}
    \label{fig:capacity_sparse}
\end{figure}

Fig. \ref{fig:capacity_sparse} demonstrates the ergodic SE of the SS scenario as a function of the transmit power per antenna, i.e., $P/(N_\rmT\iota_n^2)$, across three different system setups.
In this analysis, the parameters of $L_\rmR$ and $L_\rmT$ clusters are randomly generated, with $d _{\ell _\rmR}\sim \calU(200\lambda,400\lambda)$, $d _{\ell _\rmT} \sim \calU (50\lambda,100\lambda)$, satisfying the near-field assumption, and $\theta_{\ell _\rmR},\ \theta_{\ell _\rmT} \sim \calU(-\pi/3,\pi/3)$. 
The matrix $\bA$ is set as a diagonal matrix, with diagonal elements having amplitude $1/L_{\sparse}$ and random phases.
The correlation matrices $\cov_{\bullet,\ell_\bullet}$, $\ell_\bullet=1,\ldots,L_\bullet$ are designed to exhibit the rank-1 property, simulating the case where $\angwid \rightarrow 0$. 
The approximation in \eqref{eq:capacityapp_sparse}, the upper bound in \eqref{eq:capacitysparseub}, and the Monte-Carlo results are presented.
The congruence between the approximation and Monte-Carlo simulations is remarkable, confirming the effectiveness of our analytical approach through both the approximation and the upper bound.

The analysis, grounded in \eqref{eq:capacityapp_sparse} and \eqref{eq:capacitysparseub}, reveals that ergodic SE is influenced by several key factors: the number of multipaths, the power associated with each path, the maximum eigenvalues of the correlation matrices, and the orthogonality among the major eigenvectors of spatial correlation matrices.
The last two terms are reflected by $\chi_{\bullet, \ell}$, and $\lambda _{\bullet,\ell}$, $\ell=1,\ldots,L_{\sparse}$, respectively, which also represent the intercluster and intracluster characteristics, respectively.
These elements together signify the impact of the near-field eigenmode on system performance. The analysis suggests that the presence of more multipaths can enhance ergodic SE, even if the total multipath power remains constant, underscoring the benefits of multi-stream transmission.

In scenarios with minimal angular spread for a cluster, we observe that $\chi_{\bullet,\ell}\approx \Tr(\cov _{\bullet, \ell})$, and $\bu_{\bullet,\ell} \propto \bb_{\bullet}(\Upsilon_{\bullet,\ell})$, for $\ell=1,\ldots,L_{\sparse}$. 
This preliminarily implies that ergodic SE benefits from the deployment of additional antennas, whilst further investigation can be referred to Section \ref{sec:analysis}. Additionally, akin to far-field conditions, the asymptotic orthogonality between steering vectors of different clusters ($\ell_1 \neq \ell_2$) is approached as the number of antennas ($N_{\rmR}$ or $N_{\rmT}$) increases indefinitely:
\begin{equation}
\begin{aligned}
    &\lim _{N_\bullet\rightarrow \infty} \big| \bb _\bullet ^{\ctrans} (\Upsilon_{\ell_1}) \bb _\bullet (\Upsilon_{\ell_2}) \big| \\
    \approx &\lim _{N_\bullet\rightarrow \infty} \int _{-\frac{N_\bullet}{2}} ^ {\frac{N_\bullet}{2}} \frac{ \Big|e^{\jmath \kappa \left( D_{n_\bullet} (\Upsilon _{\ell_1}) - D_{n_\bullet} (\Upsilon _{ \ell_2}) \right)} \Big|}{ c ^{(n_\bullet)} _{\ell _1} c ^{(n_\bullet)} _{\ell _2}} \rmd n_\bullet\\
    \overset{(a)}{=} &\lim _{N_\bullet\rightarrow \infty} \frac{1}{{ c ^{(\tilde{n})} _{\ell _1} c ^{(\tilde{n})} _{\ell _2}}} \int _{-\frac{N_\bullet}{2}} ^ {\frac{N_\bullet}{2}} \Big | e^{\jmath \kappa \left( D_{n_\bullet} (\Upsilon _{\bullet, \ell_1}) - D_{n_\bullet} (\Upsilon _{\bullet, \ell_2}) \right)} \Big | \rmd n _\bullet\\
    \overset{(b)}{\rightarrow}& ~~0, ~\text{for}~\Upsilon _{\bullet,\ell_1}\neq \Upsilon _{\bullet,\ell_2}, 
\end{aligned}
\end{equation}
where (a) is due to the first mean value theorem with $-N_\bullet/2<\tilde{n}<N_\bullet/2$, and (b) is based on the conclusions in \cite{LDMA}.
Hence, the advantageous conditions for multiplexing favorable in far-field scenarios can be extended to near-field conditions as well, enhancing the system's SE.

\subsubsection{Ergodic SE in DS Scenario}

In this section, we examine the ergodic SE of the DS scenario based on the channel approximation presented in \eqref{eq:doublescatter}. For simplicity, we consider the case where $L_{\RS}=L_\rmR = L _\rmT$. The upper bound of ergodic SE is derived in the following proposition.

\begin{proposition}
\label{prop:capacityrich}
For a mid-band XL-MIMO system under the DS scenario with equal power allocation, the ergodic SE is upper bounded by  
\begin{multline}
\label{eq:capacityrich} 
C _{\RS} ^{\mathrm{ub}} \! = \! \log_2 \Bigg\{ \sum_{k=0}^{r} (k!)^2 \gamma  \sum_{\hat{\alpha}_{k}^{1}} \sum_{\hat{\alpha}_{k}^{2}} \sum_{\hat{\alpha}_{k}^{3}} \det(\bar{\bLambda}_{\rmR})_{\hat{\alpha}_{k}^{1}}^{\hat{\alpha}_{k}^{1}}   \\
\times \det(\bar{\bLambda}_{\sfC}) _{\hat{\alpha}_{k}^{2}} ^{\hat{\alpha}_{k}^{2}} \times \det(\bar{\bLambda}_{\rmT}) _{\hat{\alpha}_{k}^{3}}^{\hat{\alpha}_{k}^{3}} \Bigg\} , 
\end{multline}
where $\bar{\bLambda}_{\rmR}\in \mathbb{C}^{N_\rmR \times N_\rmR}$, $\bar{\bLambda}_\rmT \in \bbC^{N_\rmT \times N_\rmT}$ and $\bar{\bLambda}_{\sfC} \in \bbC ^ {L_{\RS}\times L_{\RS}}$ are the diagonal matrices containing the eigenvalues of $\bar{\cov}_{\rmR}$, $\bar{\cov}_{\rmT}$ and $\bA ^{\ctrans} \bA$, respectively, 
$\det(\bM) _{\hat{\alpha}_{k}^{j}} ^{\hat{\alpha}_{k}^{i}}$ denotes the determinant of a sub-matrix of $\bM$ obtained by selecting the rows and columns of $\bM$, indexed by $\hat{\alpha}_{k}^{i}$ and $\hat{\alpha}_{k}^{j}$, 
$\hat{\alpha}_k^1$, $\hat{\alpha}_k^2$, and $\hat{\alpha}_k^3$ represent all possible ordered length-$k$ subsets of the numbers $\{1,\ldots,N_\rmR\}$, $\{1,\ldots,L_{\RS}\}$, and $\{1,\ldots,N_\rmT\}$, respectively, $r = \min \{N_{\rmR},N_{\rmT},L_{\RS}\}$.
\end{proposition}
\begin{IEEEproof}
    See Appendix \ref{proof:capacityrich}.
\end{IEEEproof}

\begin{figure}[!t]
    \centering
    \includegraphics[scale=0.55]{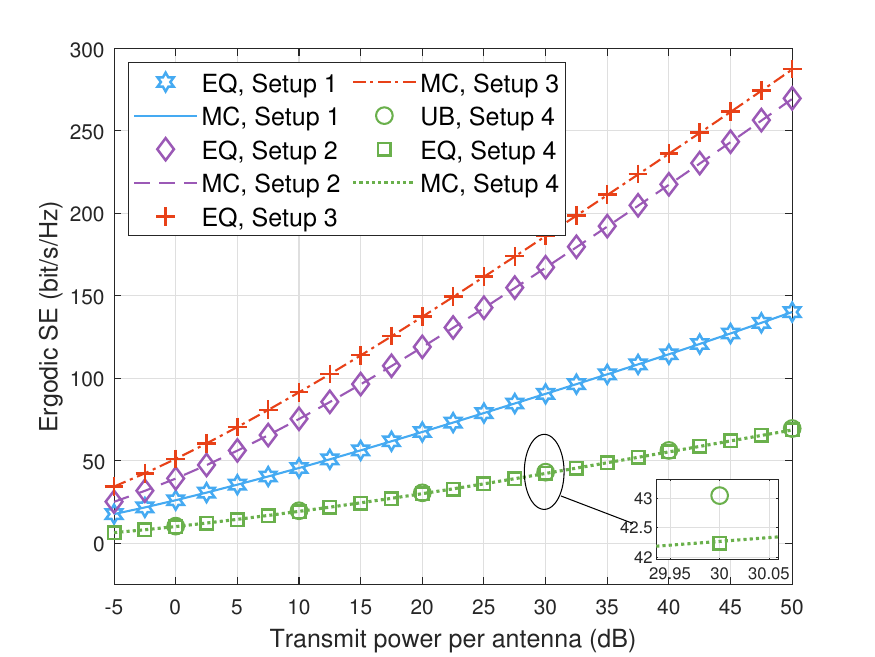}%0.5
    \caption{Ergodic SE of DS scenario against the transmit SNR. Setup 1: $N_\rmR = 256$, $N_\rmT=64$, $L_{\RS}=8$, Setup 2: $N_\rmR = 256$, $N_\rmT=64$, $L_{\RS}=16$, Setup 3: $N_\rmR = 512$, $N_\rmT=64$, $L_{\RS}=16$, Setup 4: $N_\rmR = 64$, $N_\rmT=8$, $L_{\RS}=4$.}
    \label{fig:capacity_rich}
\end{figure}

Fig. \ref{fig:capacity_rich} compares the ergodic SE of the DS scenario against the transmit power per antenna $P/(N_\rmT\iota_n^2)$ under various system configurations. 
In this analysis, the parameters of $L_\rmR$,  $L_\rmT$ clusters, $d _{\ell _\rmR}$, $d _{\ell _\rmT}$, $\theta_{\ell _\rmR}$, and $\theta_{\ell _\rmT}$ are generated in the same way as those in Fig. \ref{fig:capacity_sparse}, satisfying the near-field assumption. The angular spread for each correlation matrix is uniformly set, i.e., $\angwid=1/100$. The matrix $\bA$ is generated by $\bA _w \bA _{w} ^{\ctrans}$ and then normalized, where $\bA _{w}$ is a random matrix with entries satisfying $\CN(0,1)$.
This comparison demonstrates that the Monte-Carlo simulation results of the proposed channel model closely align with those of the equivalent model in \eqref{eq:doublescatter} under DS scenarios, thereby validating \textit{Proposition} \ref{prop:richscatter}. Moreover, the upper bound closely approximates the Monte-Carlo results, making it a rational approach to analyze the ergodic SE based on the upper bound.\footnote{Note that due to the high computational complexity of calculating the upper bound, results are presented only for cases where $N_\rmT$, $N_\rmR$ and $L _{\RS}$ are small, although the upper bound remains precise for larger parameter values.}

From \eqref{eq:capacityrich}, it is discernible that the ergodic SE is significantly influenced by the eigenvalues of $\bA^{\ctrans} \bA$, $\bar{\cov}_{\rmR}$, and $\bar{\cov}_{\rmT}$.  On one hand, the coupling relationships between clusters at the Tx and Rx sides are encapsulated by $\bar{\bLambda}_{\sfC}$, i.e., $\bA^{\ctrans} \bA$.
The structure and rank of the channel are primarily determined by $\bA^{\ctrans} \bA$, especially when the angular spread is relatively large, indicating that the ranks of $\cov_{\bullet}$, are high.
On the other hand, regarding the eigenvalue characteristics of spatial correlation matrices, under the far-field assumption, all $\cov_{\bullet,\ell_{\bullet}}, \ell_{\bullet}=1,\ldots,L_{\bullet}$, can share the same DFT eigenbases. In contrast, under the near-field assumption, deriving a closed-form expression becomes challenging; thus, we resort to numerical integration for results, which is specified in Section \ref{sec:analysis}. 

%\vspace{-0.3cm}

\subsection{OP of MIMO-MRC System Analysis}

\subsubsection{OP in SS Scenario}
Under the SS scenario, we derive an approximation for the OP under MIMO-MRC. We denote that $L_{\sparse}=\rank(\bH_{\sparse})$.
\begin{theorem}
\label{theorem:outsparse}
    For a mid-band XL-MIMO system with perfect CSI and optimal beamforming, the OP under the SS scenario is approximated by
    \begin{equation}
        P_{\out,\sparse} ^{\app} = \prod_{\ell=1}^{L_{\sparse}} \left( 1 - 2\sqrt{\frac{\gamma_{\threshold}}{\bar{\gamma} \varpi _{\ell} }} K_1\left(2\sqrt{\frac{\gamma_{\threshold}}{\bar{\gamma} \varpi _{\ell} }} \right)\right),
        \label{eq:PSSapp}
    \end{equation}
    where $\varpi _{\ell} = |\alpha_\ell|^2 \chi_{\rmR,\ell} \chi_{\rmT,\ell} \lambda_{\rmR,\ell} \lambda_{\rmT,\ell}$.
\end{theorem}
\begin{IEEEproof}
    See Appendix \ref{proof:outsparse}.
\end{IEEEproof}

\begin{figure}[!t]
    \centering
    \includegraphics[scale=0.55]{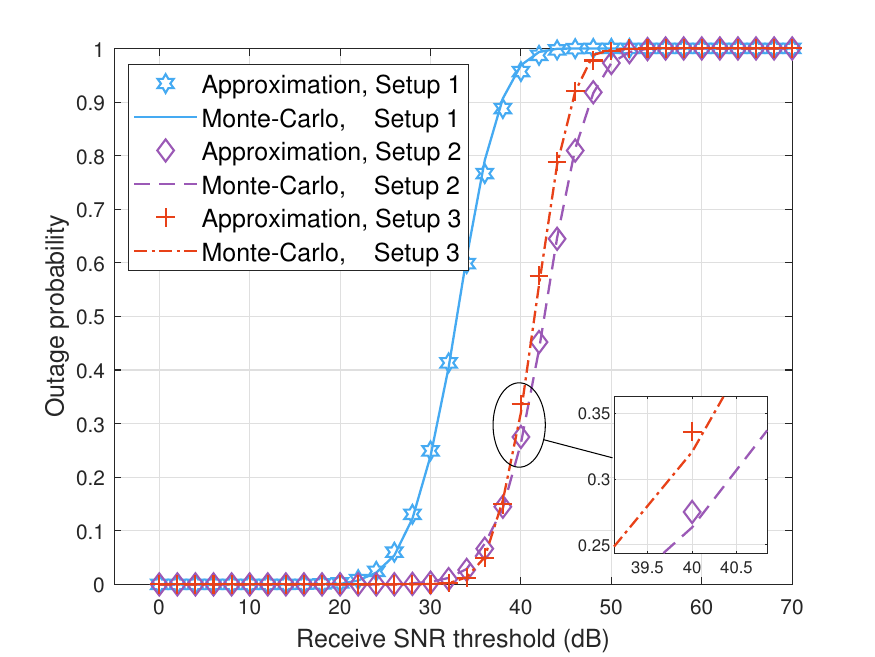} %0.45
    \caption{OP of SS scenario versus the receive SNR threshold. Setup 1: $N_\rmR = 256$, $N_\rmT=16$, $L_{\sparse}=4$, Setup 2: $N_\rmR = 512$, $N_\rmT=64$, $L_{\sparse}=4$, Setup 3: $N_\rmR = 512$, $N_\rmT=64$, $L_{\sparse}=8$.}%The transmit SNR is 0dB.
    \label{fig:outage_sparse}
\end{figure}
Fig. \ref{fig:outage_sparse} showcases the OP against the receive SNR threshold for different system configurations, with the transmit power fixed at $P/\iota_n^2=0\ \text{dB}$. 
The simulation setups are the same as those in Fig. \ref{fig:capacity_sparse}.
It is evident that the approximation results align closely with the Monte-Carlo simulations, affirming the validity of using the approximation expression for OP analysis. Notably, an increase in the number of multipaths leads to a reduction in outage, attributed to the diminished power per multipath.
Building upon previous analysis, we posit that $\chi_{\rmR,\ell} \approx \Tr(\cov _{\rmR, \ell})$ and $\chi_{\rmT,\ell} \approx \Tr(\cov _{\rmT, \ell})$, for $\ell=1,\ldots,L_{\sparse}$. Given the monotonic increase of the function $1-2\sqrt{x}K_1(2\sqrt{x})$ when $x>0$, an augmentation in the number of antennas contributes to a lower OP, thereby enhancing the quality of transmission.
From another perspective, integrating near-field characteristics into the analysis reveals that more multipaths become nearly orthogonal at both the Tx and Rx as the antenna count substantially increases. 
However, this orthogonality is not conducive to the OP of MIMO-MRC systems, due to the nearly equal $\lambda_{\bullet,\ell}$, $\ell=1, \ldots, L_{\sparse}$.
  
\subsubsection{OP in DS Scenario}

Similarly, we assume that $L_{\RS}=L_\rmR=L_\rmT$. In the context of the DS scenario, the term $\bH_{\RS} ^{\ctrans} \bH_{\RS}$ is conceptualized as a nested structure of two quadratic forms. 
Drawing on \eqref{eq:capacityrich1}, we denote $\bSigma^{\ctrans} \bSigma = \bLambda _{\sfC} \tilde{\bH}_{w,\rmT} ^{\ctrans} \bLambda_{\rmT} \tilde{\bH} _{w,\rmT} \bLambda_{\sfC} ^{\ctrans}$, the CDF of the maximum eigenvalue of $\bH^{\ctrans} _{\RS} \bH _{\RS}$ conditioned on $\bSigma^{\ctrans} \bSigma$, denoted as $F_{\lambda_{\max}}(x|\bSigma)$, can be derived based on \cite{McKay}. The distribution of eigenvalues of $\bSigma^{\ctrans} \bSigma$, denoted as $f(\blambda _{\bSigma})$, is ascertainable through the eigenvalue distribution of a quadratic form on complex Gaussian random matrices \cite{quadratic}, where $\blambda _{\bSigma}= [\lambda_{\bSigma,1},\ldots,\lambda_{\bSigma, L_{\RS}}] ^{\top}$, and $\lambda_{\bSigma,\ell}$ represents the $\ell$-th largest eigenvalue of $\bSigma$. Subsequently, the CDF of the maximum eigenvalue of $\bH^{\ctrans} _{\RS} \bH _{\RS}$ is determined through the integral
\begin{equation}
    F_{\lambda_{\max}} (x) = \int_{\mathcal{D}} F_{\lambda_{\max}} (x|\bSigma) f(\blambda_{\bSigma}) \rmd \blambda_{\bSigma},
    \label{eq:FmaxDS}
\end{equation}
where integrations occur over the domain $\mathcal{D} = \{ \infty > \lambda_{\bSigma,1} > \dots > \lambda_{\bSigma,L_{\RS}} \geq 0 \}$.
However, deriving a closed-form expression for this integral proves challenging, and the complexity of the expression also complicates direct performance analysis. Consequently, we resort to analyzing the OP using Monte-Carlo simulations for empirical insight. 

\begin{figure}[!t]
    \centering
    \includegraphics[scale=0.55]{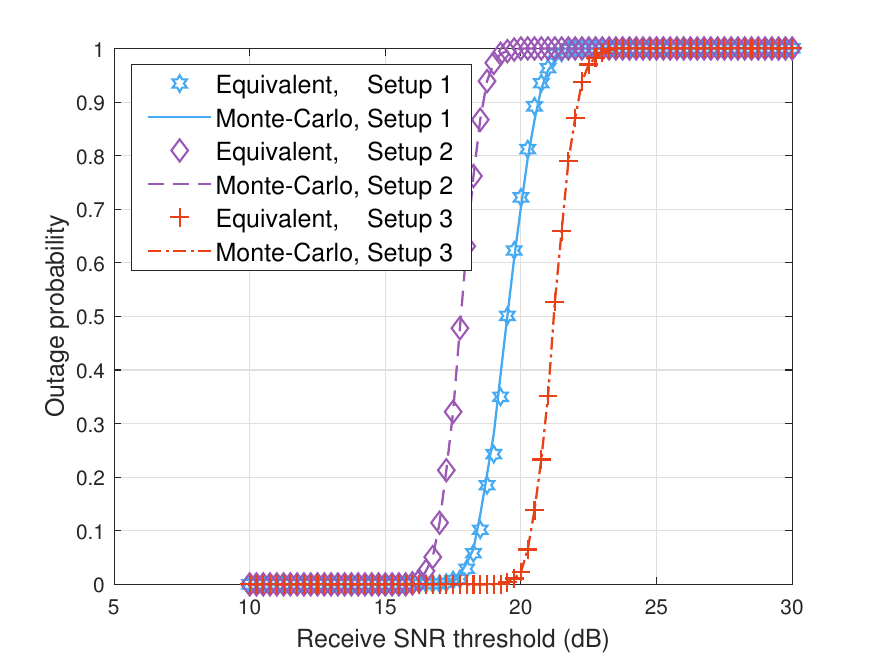} % 0.45
    \caption{OP of DS scenario versus the receive SNR threshold. Setup 1: $N_\rmR = 256$, $N_\rmT=64$, $L_{\RS}=8$, Setup 2: $N_\rmR = 256$, $N_\rmT=64$, $L_{\RS}=16$, Setup 3: $N_\rmR = 512$, $N_\rmT=64$, $L_{\RS}=16$.}
    \label{fig:outage_rich}
\end{figure}

Fig. \ref{fig:outage_rich} illustrates the OP against the receive SNR threshold for various system configurations, with the transmit power set to $P/\iota_n^2=-20\ \text{dB}$.  
The simulation settings are the same as those in Fig. \ref{fig:capacity_rich}. 
The Monte-Carlo simulations, based on the proposed channel model for the DS scenario and the equivalent expression in \eqref{eq:doublescatter}, yield similar results. This similarity further validates the accuracy of \textit{Proposition} \ref{prop:richscatter}.
Additionally, it can be inferred that an increase in the number of antennas significantly enhances the quality of the transmission link. This is because the maximum eigenvalue of $\bH_{\RS} ^{\ctrans} \bH_{\RS}$ closely correlates with the eigenvalues of $\bar{\cov} _{\rmR}$ and $\bar{\cov} _{\rmT}$ from \eqref{eq:FmaxDS}, as indicated in \eqref{eq:FmaxDS}, the characteristics of which are depicted in Section \ref{sec:analysis}. Similarly, the presence of more multipaths tends to disperse the channel power, thereby reducing the OP of a MIMO-MRC system.

\subsection{Analysis and Insights}
\label{sec:analysis}

In the analyses discussed in \eqref{eq:capacityapp_sparse}, \eqref{eq:capacitysparseub}, \eqref{eq:capacityrich} and \eqref{eq:PSSapp}, it is evident that the performance of mid-band XL-MIMO systems is significantly influenced by the \textit{eigenvalue characteristics of the spatial correlation matrices}. Therefore, this subsection focuses on the eigenvalues of these matrices, particularly in terms of their trace and distribution.

\subsubsection{Trace of Spatial Correlation Matrix}
In the SS scenario, as the spatial correlation matrices approach rank-1, the maximum eigenvalue $\chi _{\bullet, \ell _\bullet}$ in \eqref{eq:capacityapp_sparse}, \eqref{eq:capacitysparseub} and \eqref{eq:PSSapp} is approximately equal to $\Tr(\cov _{\bullet, \ell _\bullet})$. The system performance is associated with the traces of the spatial correlation matrices. Therefore, we first present the following proposition. 
 
\begin{proposition}
    The trace of the near-field spatial correlation matrix, $\cov _{\bullet, \ell _\bullet}$, derived from  \eqref{eq:NFCor}, is expressed as follows 
    \begin{align} 
        \Tr(\cov _{\bullet, \ell _\bullet}) & = \sum _{n _\bullet = 1} ^{N _\bullet} \int   \frac{1}{c _{\ell _\bullet} ^{(n _\bullet)} {}^2}   f(\Upsilon _{\ell _\bullet}) \rmd \Upsilon _{\ell _\bullet} \notag \\
        &= \int \Delta _{\bullet, \ell _\bullet} f(\Upsilon _{\ell _\bullet}) \rmd \Upsilon _{\ell _\bullet}, \label{eq:tracecov}
    \end{align} 
    where $\Delta _{\bullet, \ell _\bullet} = \bb^{\ctrans} _{\bullet} (\Upsilon _{\ell _\bullet}) \bb _{\bullet} (\Upsilon _{\ell _\bullet})$, approximated by 
    \begin{subequations}
    \begin{align}
        \Delta _{\bullet, \ell _\bullet} &\approx \frac{d _{\ell _\bullet}}{d_\rmA \cos \theta _{\ell _\bullet}} \left[ \arctan(I_1) + \arctan(I_2) \right]\\
        & = \frac{d _{\ell _\bullet}}{d_\rmA \cos \theta _{\ell _\bullet}} \left[ \arctan(I_3) + \arctan(I_4) \right],
    \end{align}
    \label{eq:Delta}
    \end{subequations}
    with the variables defined as
    \begin{subequations}
    \label{eq:I1234}
    \begin{align}  
    I_1 & = \frac{{N_{\bullet} d_\rmA} }{2d _{\ell _\bullet} \cos \theta _{\ell _\bullet}} - \tan \theta _{\ell _\bullet},\\
    I_2 & = \frac{{N_{\bullet} d_\rmA} }{2d _{\ell _\bullet} \cos \theta _{\ell _\bullet}} + \tan \theta _{\ell _\bullet},\\
    I_3 & = \frac{N_\bullet d_\rmA \cos \theta _{\ell _\bullet}}{2 d _{\ell _\bullet} - N_\bullet d_\rmA \sin \theta _{\ell _\bullet}},\\
    I_4 & = \frac{N_\bullet d_\rmA \cos \theta _{\ell _\bullet}}{2 d _{\ell _\bullet} + N_\bullet d_\rmA \sin \theta _{\ell _\bullet}},
    \end{align}
    \end{subequations}
    where $d_\rmA$ represents the spacing between adjacent array elements. Furthermore, as $\varrho \rightarrow 0$, it follows that $\chi _{\bullet, \ell _\bullet} \approx \Tr(\cov _{\bullet, \ell _\bullet}) \approx \Delta _{\bullet, \ell _\bullet}$.
    Additionally, under the assumption $d _{\ell _\bullet} \gg N _{\bullet} d_\rmA $, it can be approximated that $\Delta _{\bullet, \ell} \approx N _{\bullet}$.
\end{proposition}
\begin{IEEEproof}   
    Exchange the order of summing $n _\bullet$ and integrating $\Upsilon _{\ell _\bullet}$, we obtain the expression in \eqref{eq:tracecov}, whilst \eqref{eq:Delta} and \eqref{eq:I1234} are derived from \cite{LuICC}.
\end{IEEEproof}

\begin{figure}[!t]
    \centering
    \includegraphics[scale=0.5]{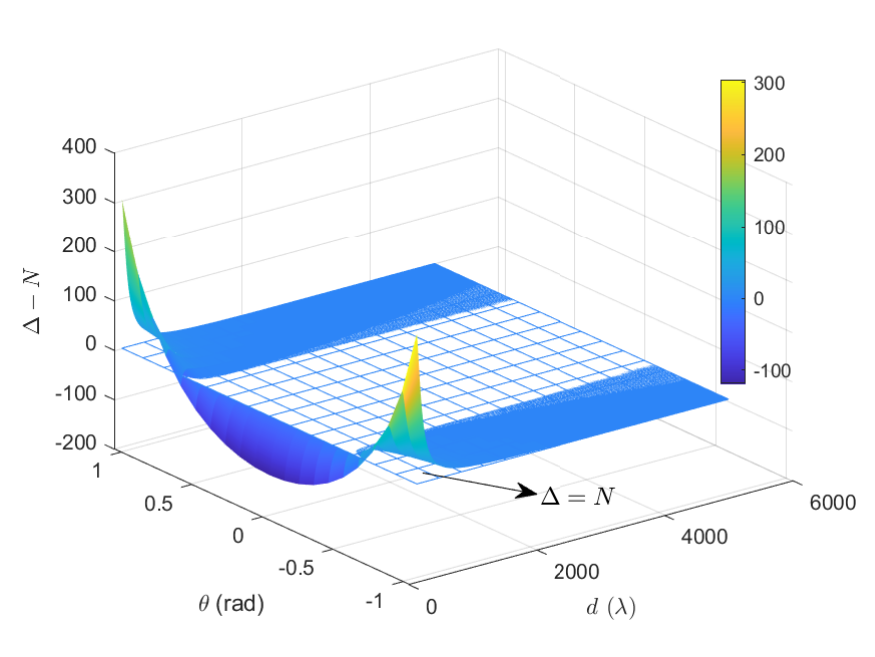}
    \caption{$\Delta - N$ versus distances and angles. $N=512$, $d\in[120\lambda,5000\lambda]$, $\theta \in [-\frac{\pi}{3},\frac{\pi}{3}]$.}
    \label{fig:cov_trace}
\end{figure}

We first focus on the characteristics of $\Delta _{\bullet, \ell}$. To capture the characteristics brought about by the near-field modeling, the term $\Delta - N$ versus the distances and angles is illustrated in Fig. \ref{fig:cov_trace}  \footnote{Note that only distances and angles in a rational range are considered, whilst extreme cases such as extremely small $d$ are left for future works.}. 
For simplicity, the subscripts $\bullet$ and $\ell$ are omitted.
It can be observed that when $d$ is large, $\Delta \approx N$ is satisfied for different angles, because the near-field characteristics are equivalent to the far-field characteristics at a large $d$ and non-stationarities can be omitted.
However, when $d$ is small, the term $\Delta- N$ first decreases to less than 0, then increases to greater than 0 with the increase of $\theta$.
We denote the two zeros as $\theta _1(d)$ and $\theta _2 (d)$, satisfying $\theta _1(d) < 0 < \theta _2(d)$, which are functions of $d$.
Based on the characteristics of $\Delta$, we next consider the influence on system performance.

Our consideration can be divided into two aspects: $\Delta-N >0$ and $\Delta - N <0$, which correspond further to two angular regions: $\vartheta_1=\{\theta|-\frac{\pi}{2}<\theta < \theta_1(d),\theta_2(d)<\theta <\frac{\pi}{2} \}$ and $\vartheta_2=\{\theta|\theta_1(d) < \theta < \theta_2(d)$\}.
In region $\vartheta_1$, we have 
\begin{equation}
    \chi _{\bullet,\ell _\bullet}\! \approx \! \int \Delta _{\bullet, \ell _\bullet} f(\Upsilon _{\ell _\bullet}) \rmd \Upsilon _{\ell _\bullet} \! \geq \! \int N _{\bullet} f(\Upsilon _{\ell _\bullet}) \rmd \Upsilon _{\ell _\bullet}\!=\!N _{\bullet},
\end{equation}
due to the property $\int f(\Upsilon _{\ell _\bullet}) \rmd \Upsilon _{\ell _\bullet}=1$.
Whilst in region $\vartheta_2$, $\chi _{\bullet,\ell _\bullet}\leq N _{\bullet}$ is inferred.
This indicates that system performance also suffers from a \textit{non-stationarity}, i.e., clusters in different regions have non-stationary contributions.
Afterward, we focus on the increase of array elements. It can be obtained from \eqref{eq:Delta} that $\chi _{\bullet, \ell}$ in \eqref{eq:capacityapp_sparse}, \eqref{eq:capacitysparseub} and \eqref{eq:PSSapp} increases with the increase of $N_\bullet$. However, when $N _\bullet \rightarrow \infty$, $\chi _{\bullet, \ell}$ tends to be constant with $\frac{\pi d _{\ell _\bullet}}{d_\rmA \cos \theta _{\ell _\bullet}}$, which is in accordance with \cite{LuICC}.
Therefore, we can derive the following key insight. 
  
\textit{Key Insight 1}: In the SS scenario, the performance of mid-band XL-MIMO systems, including metrics such as ergodic SE and MIMO-MRC OP, benefits from an increase in the number of array elements. However, the performance gains tend to plateau as the number of array elements approaches infinity, a behavior that contrasts with massive MIMO systems under the far-field assumption. Therefore, it is advisable to moderately increase the number of array elements, taking into account the specific conditions of actual deployment.

\begin{figure}[!t]
    \centering
    \includegraphics[scale=0.55]{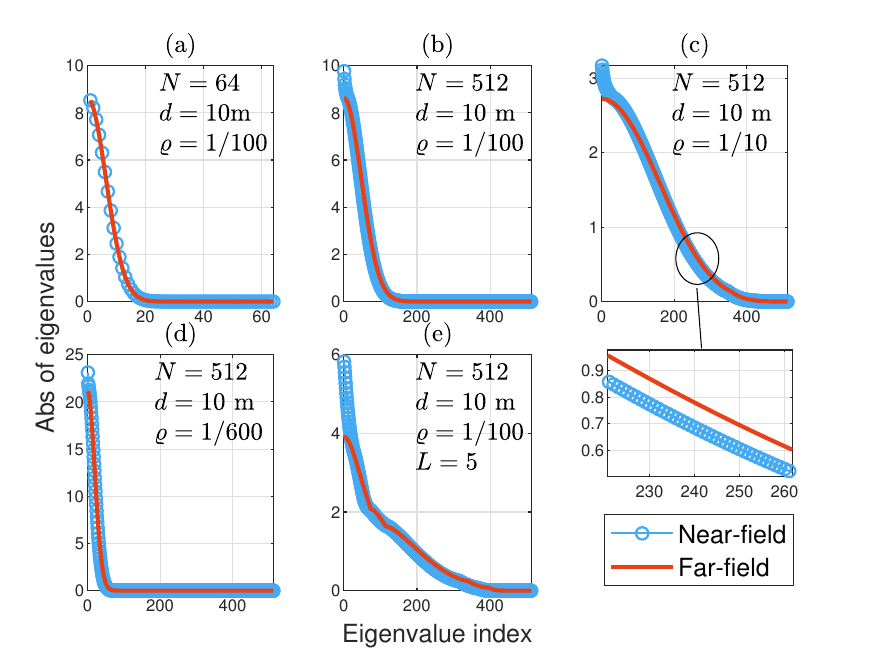} %0.45
    \caption{Comparisons of eigenvalues of spatial correlation matrices under different configurations and region $\vartheta _1$. (a), (b), (c), and (d) present the eigenvalues of a single correlation matrix, (e) describes the eigenvalues of the mean of 5 spatial correlation matrices. The mean angles are set to $\pi/8$.} 
    \label{fig:CorEigCmp}
\end{figure}

\begin{figure}[!t]
    \centering
    \includegraphics[scale=0.55]{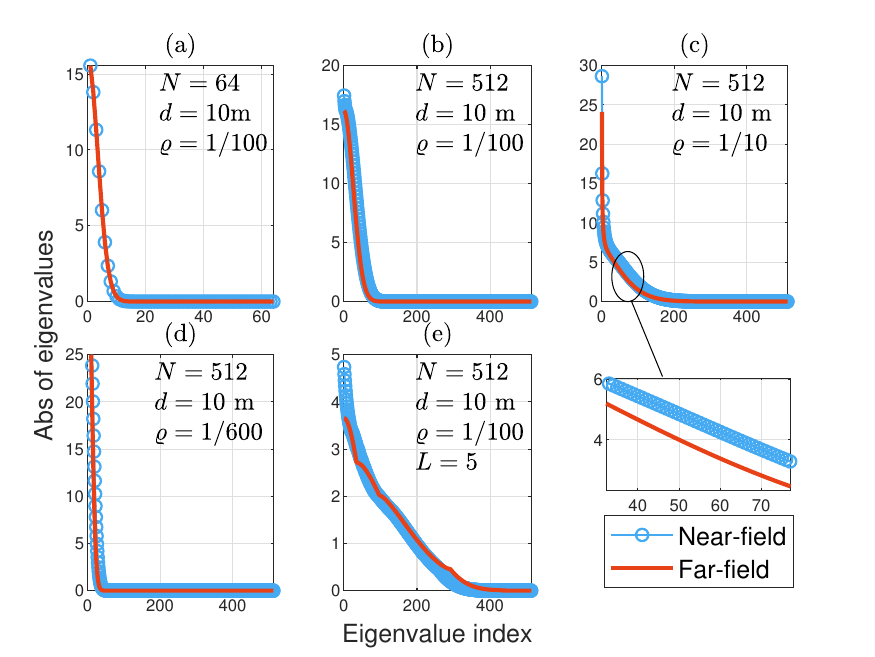} %0.45
    \caption{Comparisons of eigenvalues of spatial correlation matrices under different configurations and region $\vartheta _2$. The descriptions of (a), (b), (c), (d), and (e) are the same as Fig. \ref{fig:CorEigCmp}. The mean angles are set to $\pi/3$.} 
    \label{fig:CorEigCmp1}
\end{figure}

\subsubsection{Distribution of Eigenvalues of Spatial Correlation Matrix}
In the DS scenario, the system performance in \eqref{eq:capacityrich} and \eqref{eq:FmaxDS} is related to the distribution of eigenvalues.
Similarly, our consideration can be divided into two aspects: region $\vartheta _1$ and region $\vartheta _2$.
Since it is challenging to derive closed-form expressions for the spatial correlation matrices, we present numerical results from the two perspectives above.

As illustrated in Fig. \ref{fig:CorEigCmp} and Fig. \ref{fig:CorEigCmp1}, we compare the eigenvalues of spatial correlation matrices between the far-field and near-field conditions under various parameters. Fig. \ref{fig:CorEigCmp} and Fig. \ref{fig:CorEigCmp1} correspond to $\Delta-N <0$ and $\Delta-N >0$, respectively. 
In these figures, $N$ is the number of antennas, $d$ is the distance coordinate of cluster center, $\angwid$ reflects the angular spread.
The mean angles are set as $\pi/8$ and $\pi/3$ for Fig. \ref{fig:CorEigCmp} and Fig. \ref{fig:CorEigCmp1}, respectively. The mean angles of $L=5$ clusters in both figures are generated randomly from $-\pi/3$ to $\pi/3$.

Observing Fig. \ref{fig:CorEigCmp}(a), Fig. \ref{fig:CorEigCmp}(b), Fig. \ref{fig:CorEigCmp1}(a) and \ref{fig:CorEigCmp1}(b), it is evident that increasing the array's dimension enhances the range of the near field, which further causes the eigenvalues of the spatial correlation matrix under near-field conditions to surpass those under far-field conditions. According to Fig. \ref{fig:CorEigCmp}(c), the \textit{dominant} eigenvalues of the spatial correlation matrix under near-field conditions surpass those under far-field conditions for both large and small $\angwid$ alike.  
However, the \textit{overall}  eigenvalues under near-field conditions surpass those under far-field conditions, as reflected by Fig. \ref{fig:CorEigCmp1}(c).

As anticipated, with a small $\angwid$, both near-field and far-field correlation matrices tend toward a rank-1 distribution, showing similar eigenvalue distributions. Additionally, when considering multiple clusters, as depicted in Fig. \ref{fig:CorEigCmp}(e) and Fig. \ref{fig:CorEigCmp1}(e), we take the average of $L=5$ spatial correlation matrices as an example. Evidently, the eigenvalues under near-field modeling exceed those modeled under far-field conditions. The performance of multiple spatial correlation matrices can be explained by the additivity of non-overlapping multiple angular distributions.

Based on the observations above, the overall eigenvalues of spatial correlation matrices of near-field clusters with significant deviation from the normal direction tend to be larger than those under the far-field assumption. While only the dominant eigenvalues under near-field modeling are larger than those under far-field modeling when the incident angle is close to the normal direction. 
From the system performance perspective, when the overall eigenvalues under near-field modeling exceed those under far-field modeling, the enhancement in SE can be obtained through spatial multiplexing transmission. Conversely, if only the dominant eigenvalues under near-field modeling are larger than those under far-field modeling, the enhancement can be reflected through beamforming with several data streams. Consequently, these characteristics lead to the following key insight.

\textit{Key Insight 2}: In the DS scenario, the advantages of mid-band XL-MIMO systems in SE can be reflected through multiplexing and beamforming, which is determined by the positions of near-field clusters.

In the forthcoming section, we will delve into the frequency-related parameters and further examine the performance of mid-band XL-MIMO systems against other system configurations. This examination will leverage both the proposed channel model and simulations grounded in practical system configurations, aiming to provide a comprehensive comparison.

%\section{Simulation Results}
\section{Performance Comparisons Under Different System Configurations}
In addition to theoretical analyses, our objective is to benchmark the performance of mid-band XL-MIMO systems against systems configured under various practical settings. The frequency-related parameters for this comparison are derived from standardized technical reports or measurement results. These parameters are integral to the proposed channel model configuration. Specifically, the path loss is modeled based on the Urban Micro (UMi) scenario as per \cite{38901}:
\begin{equation}
    \mathsf{PL} = 32.4 + 21\log_{10}\left( d_{\mathrm{TR}} \right) + 20\log_{10}(f_c)\ [\mathrm{dB}],
\end{equation}
where $d_{\mathrm{TR}}$ denotes the distance between the transmitter and receiver, and $f_c$ represents the carrier frequency. Following the findings in \cite{MiaoJSAC}, the number of clusters for different frequency bands is modeled as 
\begin{equation}
    N_{\mathrm{cluster}} = 3.41 \cdot e^{-0.17 f_c} + 1.86,
\end{equation}
with each cluster containing five rays. The cluster center positions $(d_{\ell_\bullet},\theta_{\ell_ \bullet})$ are randomly generated within specified ranges, with $d_{\ell_\bullet}\sim \mathcal{U}(10\mathrm{m},15\mathrm{m})$, $\theta_{\ell_\bullet} \sim \mathcal{U}(-\pi/3,\pi/3)$.

To contrast the performance of typical MIMO systems across different frequency bands with mid-band XL-MIMO systems, we delineate the parameters for various system configurations in Table \ref{tab:systemconfig}. 
Specifically, setup 1 and setup 3 simulate Sub-6 GHz and mmWave MIMO systems, respectively, with bandwidths of 100 MHz and 1600 MHz, which are typical bandwidth configurations for FR1 and FR2 \cite{38104}. Setup 2 simulates the mid-band XL-MIMO systems with different configurations.
The distance between the transmitter and receiver is fixed at $d_{\mathrm{TR}}=20$ m. The channel matrix $\bA$ in setups 1 and 3 are assumed to be diagonal and non-sparse, respectively, with randomly generated elements, whereas both conditions of $\bA$ are evaluated for setup 2. The analyses are underpinned by 1,000 Monte-Carlo simulations.
For throughput comparisons, we evaluate SE based on bandwidth, while reliability assessments hinge on receive SNR for intuitive insights.

\begin{figure}[!t]
    \centering
    \includegraphics[scale=0.55]{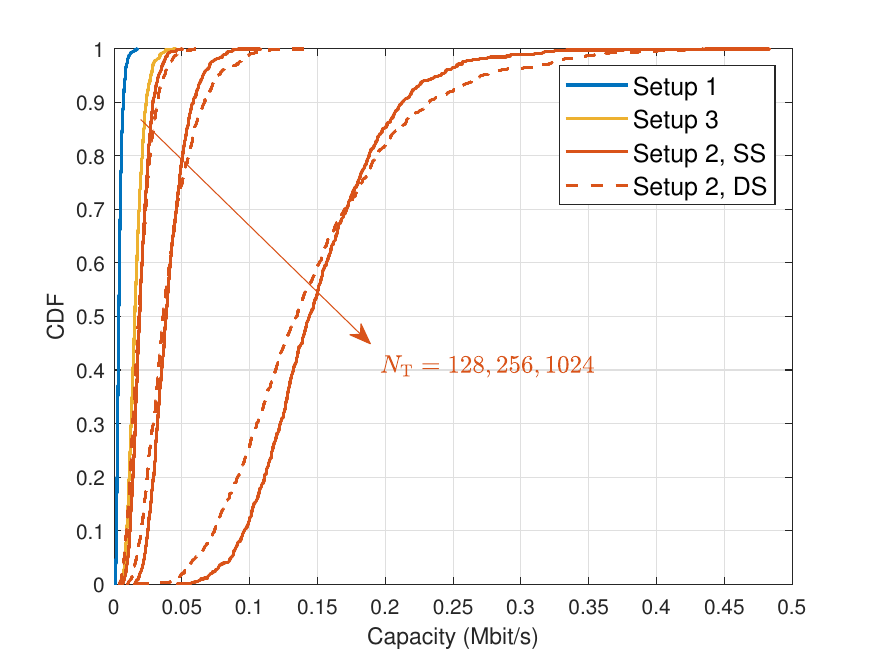} %0.45
    \caption{Comparisons of capacity under different system configurations.}
    \label{fig:capacity_system}
\end{figure}

\begin{table}[!t]
	\caption{Parameters of different system configurations}\label{tab:systemconfig}
	\centering
	\begin{tabular}{|c|c|c|c|}%|c|c|c| |c|c|c|c|
	\hline
    Parameters & Setup 1 & Setup 2 & Setup 3\\
    \hline
	$f_c$ & 3.5 GHz & 7 GHz & 28 GHz \\
    \hline
	$N_\rmT, N_\rmR$  & $32,4$ & $[128,256,1024],8$ & $512,16$ \\
    \hline
    Bandwidth & 100 MHz & 500 MHz & 1600 MHz \\
	\hline
	\end{tabular}
\end{table}

Fig. \ref{fig:capacity_system} exhibits the cumulative distribution function (CDF) of capacity across different system configurations, indicating that systems configured as setup 2 excel due to their broader bandwidth and augmented array elements. The empirical data corroborate that augmenting the number of array elements at both the base station and the user equipment effectively bolsters system capacity, underscoring the efficacy and potential of mid-band wireless communication systems in throughput enhancement.

\begin{figure}[!t]
    \centering
    \includegraphics[scale=0.55]{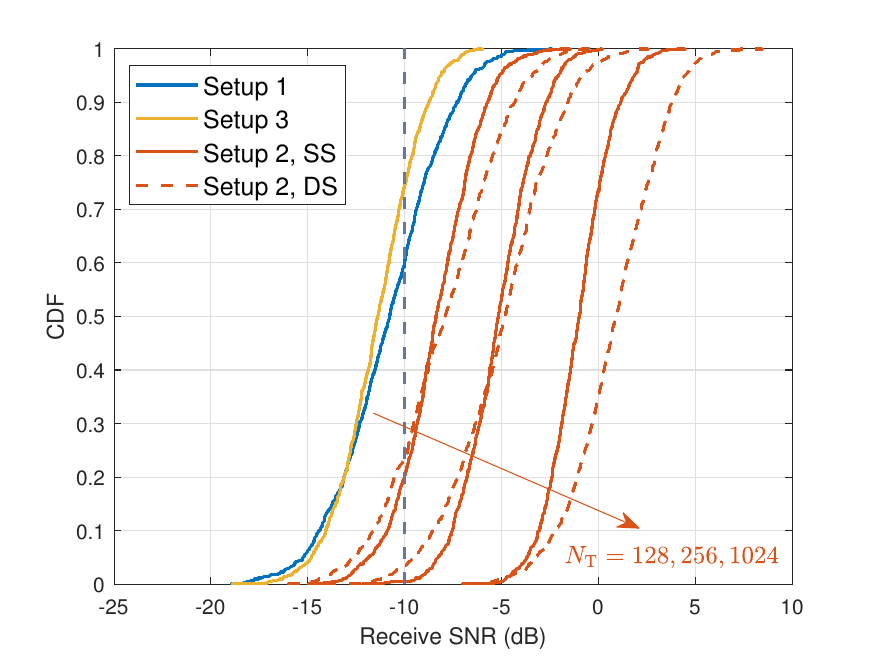} %0.45
    \caption{Comparisons of receive SNR of MIMO-MRC system under different system configurations}
    \label{fig:outage_system}
\end{figure}

Fig. \ref{fig:outage_system} showcases the CDF of receive SNR under assorted practical system configurations, assuming downlink transmission with a transmit power of 40 dB. Notably, when employing beamforming strategies at both the transmitter and receiver, mid-band MIMO systems demonstrate superior transmission link quality, particularly with an expanded antenna array. For instance, with a receive SNR threshold of -10 dB, the OP of systems configured as setup 2 approaches zero with 256 and 8 antennas at the Tx and Rx, respectively.
 
This comparative analysis reveals that, relative to Sub-6 GHz MIMO systems, mid-band XL-MIMO systems, benefitting from an increased antenna count and enhanced bandwidth, excel in both throughput and reliability despite more pronounced large-scale fading. When juxtaposed with mmWave MIMO systems, which are adversely affected by severe large-scale fading and rank deficiency, mid-band XL-MIMO systems still achieve higher transmission rates and more reliable transmission, even with a comparable antenna count at the transmitter and receiver. These simulation results further affirm the superior performance of mid-band XL-MIMO systems in terms of throughput, reliability, and network coverage, facilitated by increased array elements, channel rank enhancement, and broader transmission bandwidth.

\section{Conclusion}
The mid-band XL-MIMO system, integrating the mid-band spectrum and XL-MIMO technology, emerges as a promising facilitator for the advancement of future wireless communication systems. This development is further supported by ongoing efforts in standardization. To conduct effective performance analysis and comparisons, we introduce a novel analytical channel model grounded on an extensive analysis of potential channel characteristics specific to the mid-band spectrum. Utilizing this model, we delve into the ergodic SE and the OP of MIMO MRC systems, highlighting them as pivotal metrics. Specifically, we derive closed-form approximations or performance bounds for two illustrative scenarios, facilitating analysis and comparison of the impact exerted by the potential channel characteristics of mid-band XL-MIMO systems on performance metrics.
The influences on system performance are further analyzed from the perspective of eigenvalue characteristics of spatial correlation matrices. Simulations based on the proposed channel model and practical parameters underscore the potential and superiority of mid-band XL-MIMO systems.
The analysis and numerical results demonstrate that the performance of mid-band XL-MIMO systems benefits from the combination of increased array elements, moderate large-scale fading, enhanced channel rank, and enlarged transmission bandwidth, which are expected to exceed those of MIMO systems and massive MIMO systems under Sub-6 GHz and mmW bands.

{\appendices
\section{Preliminaries}
\label{sec:appexp}
\begin{lemma}\label{lemma:exp}
    Assuming $X$ and $Y$ are two independent random variables, satisfying exponential distribution as $X\sim \exp(\lambda_1)$ and $Y\sim \exp(\lambda_2)$ $\lambda_1, \lambda_2 >0$, respectively, then the cumulative distribution function (CDF) and probability density function (PDF) of the random variable $Z=XY$ are denoted as \eqref{eq:CDF} and \eqref{eq:PDF}, respectively.
    \begin{equation}
        F_{Z}(z) = 1 - 2\sqrt{\lambda_1 \lambda_2 z} K_{1} \left(2 \sqrt{\lambda_1 \lambda_2 z} \right),
    \label{eq:CDF}
    \end{equation}
    \begin{equation}
        f_Z(z) = 2\lambda_1 \lambda_2 K_0 \left( 2 \sqrt {\lambda_1 \lambda_2 z} \right),
    \label{eq:PDF}
    \end{equation}
    where $K_{\nu}(\cdot)$ is the modified Bessel function of the second kind with order $\nu$.
\end{lemma}
\begin{IEEEproof}
    The PDF of X and Y are $f_X(x)= \lambda_1 e^{-\lambda_1 x}, x\geq 0$ and $f_Y(y) = \lambda_2 e^{-\lambda_2 y}, y\geq 0$, then the CDF of $Z=XY$ is
    \begin{equation}
    \begin{aligned}
        F_Z(z) \!  = \! P(Z\leq z) \! & = \! \int _{0} ^{+\infty} \int _{0} ^{\frac{z}{y}} \lambda_{1} \lambda_{2}e^{-\lambda_1 x} e^{-\lambda_2 y} \rmd x \rmd y\\
        & = \! \int _{0} ^{+\infty} \lambda_2 e ^{-\lambda_2 y} (1 - e ^{- \lambda_1 \frac{z}{y}}) \rmd y,
    \end{aligned}
    \end{equation}
then apply the integral property in \cite{TableInt} and we obtain the conclusions above.
\end{IEEEproof}

\section{Proof of Proposition \ref{prop:kronecker}}
\label{proof:kronecker}
The spatial correlation coefficient for a link from cluster $\ell_\rmT$ to cluster $\ell_\rmR$ is defined as
\begin{equation}
\setlength\abovedisplayskip{5pt}
\setlength\belowdisplayskip{5pt}
\xi^{(n_{\rmR_1},n_{\rmT_1})}_{n_{\rmR_2},n_{\rmT_2}} = \bbE \left\{ \frac{h_{n_{\rmR_1},n_{\rmT_1},\ell_\rmR,\ell_\rmT} h^{*}_{n_{\rmR_2},n_{\rmT_2},\ell_\rmR,\ell_\rmT} }{|\alpha _{\ell _\rmR, \ell _\rmT}|^2} \right\}.
\label{eq:spatialcorr}
\end{equation}
Upon substituting \eqref{eq:GBSM} into \eqref{eq:spatialcorr}, we arrive at a detailed expression in \eqref{eq:spatialcorr1}, appearing at the top of the next page.
\newcounter{TempEqCnt}
\setcounter{TempEqCnt}{\value{equation}}
\begin{figure*}[!t]
\normalsize
%\hrulefill
\begin{equation}
\begin{aligned}
    &\xi^{( n_{\rmR _1},n_{\rmT _1} )} _{n_{\rmR _2},n_{\rmT _2}} = \! \frac{\sum _{q_{\rmR}=1}^{Q_{\rmR}} \sum_{q_\rmT=1}^{Q_\rmT}}{Q_{\rmR}Q_{\rmT}}
    \bbE \left\{ |g_{\ell_\rmR,q_\rmR} g_{\ell_\rmT,q_\rmT}|^2\frac{e^{\jmath \kappa \left( D_{n_{\rmR_2}}(\Upsilon_{\ell_\rmR,q_\rmR}) - D_{n_{\rmR_1}}(\Upsilon_{\ell_\rmR,q_\rmR})\right)} }{c ^{(n_{\rmR_2})} _{\ell_\rmR,q_\rmR} c ^{(n_{\rmR_1})} _{\ell_\rmR,q_\rmR}}
    \cdot \frac{e^{\jmath \kappa \left( D_{n_{\rmT_1}}(\Upsilon_{\ell_\rmT,q_\rmT}) - D_{n_{\rmT_2}}(\Upsilon_{\ell_\rmT,q_\rmT}) \right) } }{c ^{(n_{\rmT_2})} _{\ell_\rmT,q_\rmT} c ^{(n_{\rmT_1})} _{\ell_\rmT,q_\rmT}} \right\}\\
    &=\! \frac{\sum _{q_\rmR=1} ^{Q_\rmR} \bbE\{|g_{\ell_\rmR, q_{\rmR}}| ^2\}}{Q_\rmR} \frac{e^{\jmath \kappa \left( D_{n_{\rmR_2}}(\Upsilon_{\ell_\rmR,q_\rmR}) - D_{n_{\rmR_1}}(\Upsilon_{\ell_\rmR,q_\rmR})\right)} }{c ^{(n_{\rmR_2})} _{\ell_\rmR,q_\rmR} c ^{(n_{\rmR_1})} _{\ell_\rmR,q_\rmR}}  \cdot 
    \frac{\sum _{q_\rmT=1} ^{Q_\rmT} \bbE\{|g_{\ell_\rmT, q_{\rmT}}| ^2\}}{Q_\rmT} \frac{e^{\jmath \kappa \left( D_{n_{\rmT_1}}(\Upsilon_{\ell_\rmT,q_\rmT}) - D_{n_{\rmT_2}}(\Upsilon_{\ell_\rmT,q_\rmT}) \right) } }{c ^{(n_{\rmT_2})} _{\ell_\rmT,q_\rmT} c ^{(n_{\rmT_1})} _{\ell_\rmT,q_\rmT}}.
\end{aligned}
\label{eq:spatialcorr1}
\end{equation}
%\vspace*{4pt}%4pt
\hrulefill
\end{figure*}    
We assume that the link is coupled through $\alpha_{\ell_\rmR, \ell_\rmT}$ at the cluster level, implying that the statistical characteristics on the Tx and Rx sides are assumed to be independent. Consequently, \eqref{eq:spatialcorr1} can be further represented as $\xi^{(n_{\rmR_1},n_{\rmT_1})}_{n_{\rmR_2},n_{\rmT_2}} =  \xi_{n_{\rmR _2}}^{(n_{\rmR _1})} \cdot \xi_{n_{\rmT _2}}^{(n_{\rmT _1})} {}^{*}$, where the superscript $*$ represents the conjugate.  
Here, $\xi_{n_{\rmR_2}}^{(n_{\rmR_1})}$ and $\xi_{n_{\rmT_2}}^{(n_{\rmT_1})}$ are the spatial correlation coefficients at the Rx and Tx, respectively, defined as
\begin{equation}
\begin{aligned}
\xi_{n_{\bullet_2}}^{(n_{\bullet_1})}\! =\! \frac{\sum \limits_{q_\bullet=1} \limits^{Q_\bullet} \bbE\{|g_{\ell_\bullet, q_{\bullet}}| ^2\}}{Q_\bullet} \frac{e^{\jmath \kappa \left( D_{n_{\bullet_2}}(\Upsilon_{\ell_\bullet,q_\bullet}) - D_{n_{\bullet_1}}(\Upsilon_{\ell_\bullet,q_\bullet})\right)} }{c ^{(n_{\bullet_2})} _{\ell_\bullet,q_\bullet} c ^{(n_{\bullet_1})} _{\ell_\bullet,q_\bullet}},
\end{aligned}
\label{eq:xiX}
\end{equation}
for $\bullet \in \{\rmR,\rmT\}$.
Arranging $\xi_{n_{\rmR_2}}^{(n_{\rmR_1})}$ and $\xi_{n_{\rmT_2}}^{(n_{\rmT_1})}$ into matrices $\cov_{\rmR, \ell _\rmR}$ and $\cov _{\rmT, \ell _\rmT}$, satisfying $[\cov_{\rmR,\ell_\rmR}]_{n_{\rmR_1},n_{\rmR_2}} = \xi_{n_{\rmR_2}}^{(n_{\rmR_1})}$ and $[\cov_{\rmT,\ell_\rmT}]_{n_{\rmT_1},n_{\rmT_2}} = \xi_{n_{\rmT_2}}^{(n_{\rmT_1})}$, we obtain \eqref{eq:prop1} in \textit{Proposition} \ref{prop:kronecker} due to the Herimitian characteristics of $\cov _{\bullet, \ell _\bullet}$.

%\vspace{-0.3cm}
\section{Proof of Proposition \ref{prop:eigenmode}}
\label{proof:eigenmode}
%\newcounter{TempEqCnt}
\setcounter{TempEqCnt}{\value{equation}}
\begin{figure*}[!t]
\normalsize	
%\hrulefill
\begin{equation}
\begin{aligned}
\cov_{\rmR} & = \bbE \left\{ \left( \sum_{\ell_{\rmR_1}} \sum_{\ell_{\rmT_1}} \alpha_{\ell_{\rmR_1},\ell_{\rmT_1}} \cov^{\frac{1}{2}} _{\rmR,\ell_{\rmR_1}} \bg_{\rmR,\ell_{\rmR_1}} \bg ^{\ctrans} _{\rmT,\ell_{\rmT_1}} \cov^{\frac{\ctrans}{2}} _{\rmT,\ell_{\rmT_1}} \right)  \left( \sum_{\ell_{\rmR_2}} \sum_{\ell_{\rmT_2}} \alpha ^{*} _{\ell_{\rmR_2},\ell_{\rmT_2}} \cov^{\frac{1}{2}} _{\rmT,\ell_{\rmT_2}} \bg_{\rmT,\ell_{\rmT_2}} \bg ^{\ctrans} _{\rmR,\ell_{\rmR_2}} \cov^{\frac{\ctrans}{2}} _{\rmR,\ell_{\rmR_2}} \right) \right\}\\
& = \sum_{\ell_{\rmR_1}} \sum_{\ell_{\rmT_1}} \sum_{\ell_{\rmR_2}} \sum_{\ell_{\rmT_2}} \bbE \left\{ \alpha_{\ell_{\rmR_1},\ell_{\rmT_1}} \alpha^{*} _{\ell_{\rmR_2},\ell_{\rmT_2}} \cov^{\frac{1}{2}} _{\rmR,\ell_{\rmR_1}} \bg_{\rmR,\ell_{\rmR_1}} \bg ^{\ctrans} _{\rmT,\ell_{\rmT_1}} \cov^{\frac{\ctrans}{2}} _{\rmT,\ell_{\rmT_1}} \cov^{\frac{1}{2}} _{\rmT,\ell_{\rmT_2}} \bg_{\rmT,\ell_{\rmT_2}} \bg ^{\ctrans} _{\rmR,\ell_{\rmR_2}} \cov^{\frac{\ctrans}{2}} _{\rmR,\ell_{\rmR_2}} \right\}.
\end{aligned}
\label{eq:CorR}
\end{equation}
%\vspace*{4pt}%4pt
\hrulefill
\end{figure*}

The correlation matrices at the Tx and the Rx are defined as $\cov_{\rmT}=\bbE\left\{ \bH^{\ctrans} \bH \right\}$ and $\cov_\rmR=\bbE \left\{ \bH \bH^{\ctrans} \right\}$, respectively. Here we take the Rx as an example. The receive correlation matrix is written as \eqref{eq:CorR}, shown at the top of the next page.
When $\ell_{\rmT_1} \neq \ell_{\rmT_2}$, we have
\begin{equation}
    \bbE \left\{\bg ^{\ctrans} _{\rmT,\ell_{\rmT_1}} \cov ^{\frac{\ctrans}{2}} _{\rmT,\ell_{\rmT_1}} \cov^{\frac{1}{2}} _{\rmT,\ell_{\rmT_2}} \bg_{\rmT,\ell_{\rmT_2}} \right\}=0.
\end{equation}
When $\ell_{\rmT_1} = \ell_{\rmT_2} = \ell_\rmT$ and $\ell_{\rmR_1} \neq \ell_{\rmR_2}$, the following expression also satisfies due to the independent characteristics of $\bg_{\rmR,\ell _{\rmR_1}}$ and $\bg _{\rmR,\ell_{\rmR_2}}$.
\begin{equation}
    \setlength\abovedisplayskip{5pt}
    \setlength\belowdisplayskip{5pt}
    \bbE \left\{ \bg_{\rmR,\ell_{\rmR_1}} \bg ^{\ctrans} _{\rmT,\ell_{\rmT}} \cov _{\rmT,\ell_{\rmT}} \bg_{\rmT,\ell_{\rmT}} \bg ^{\ctrans} _{\rmR,\ell_{\rmR_2}} \right\} = \mathbf{0}.
\end{equation}
Therefore, only the case $\ell_{\rmR_1}=\ell_{\rmR_2} = \ell_{\rmR}$ and $\ell_{\rmT_1}=\ell_{\rmT_2} = \ell_{\rmT}$ needs to be considered. Define $\epsilon _{\ell_\rmT} = \bg ^{\ctrans} _{\rmT,\ell_\rmT} \cov _{\rmT,\ell_\rmT} \bg _{\rmT,\ell_\rmT}$ and apply the property in \cite{matvariate}, we have
\begin{equation}
\begin{aligned}
    \bbE & \left\{ \cov^{\frac{1}{2}} _{\rmR,\ell_{\rmR}} \bg_{\rmR,\ell_{\rmR}} \epsilon _{\ell_\rmT} \bg ^{\ctrans} _{\rmR,\ell_{\rmR}} \cov^{\frac{\ctrans}{2}} _{\rmR,\ell_{\rmR}} \right\}\\
    &= \cov^{\frac{1}{2}} _{\rmR,\ell_{\rmR}}  \bbE \left\{ \bg_{\rmR,\ell_{\rmR}} \epsilon _{\ell_\rmT} \bg ^{\ctrans} _{\rmR,\ell_{\rmR}} \right\} \cov^{\frac{1}{2}} _{\rmR,\ell_{\rmR}} = \epsilon _{\ell_\rmT} \cov_{\rmR,\ell_{\rmR}}.%\Tr(\epsilon)
    \label{eq:CovR1}
\end{aligned}
\end{equation}
Then the $\epsilon _{\ell_\rmT}$ can be approximated by $\bbE\{\epsilon _{\ell_\rmT}\}$ \cite{JHWang}
\begin{equation}
    \setlength\abovedisplayskip{5pt}
    \setlength\belowdisplayskip{5pt}
    \bbE\{ \epsilon _{\ell_\rmT} \} = \bbE\{ \bg ^{\ctrans} _{\rmT,\ell_\rmT} \cov _{\rmT,\ell_\rmT} \bg _{\rmT,\ell_\rmT} \} = \Tr(\cov_{\rmT,\ell_\rmT}).
    \label{eq:CovR2}
\end{equation}
Substituting \eqref{eq:CovR1} and \eqref{eq:CovR2} into \eqref{eq:CorR} then the right hand of \eqref{eq:eigenmode} can be obtained.
Similarly, the correlation matrix at the Tx side can also be obtained.

\section{Proof of Proposition \ref{prop:richscatter}}
\label{proof:richscatter}
Since the $\bH_{\rmR}$ and $\bH_{\rmT}$ in \eqref{eq:channel} are modeled independently while coupled by $\bA$, we then analyze $\bH_{\rmR}$ and $\bH_{\rmT}$ seperately. Considering a general expression, the statistical properties of one side $\bullet$, $\bullet \in \{\rmT,\rmR \}$, satisfy
\begin{equation}
\begin{aligned}
    \bbE  \{\bH_{\bullet} \bH_{\bullet} ^{\ctrans} \} = \bbE \bigg \{ \left[\cov_{\bullet,1}^{\half} \bg_{\bullet,1},\ldots, \cov_{\bullet,L _{\bullet}} ^{\half} \bg_{\bullet,L_\bullet} \right] &\\
    \left[\cov_{\bullet,1}^{\frac{1}{2}} \bg _{\bullet,1},\ldots, \cov_{\bullet,L _{\bullet}} ^{\frac{1}{2}} \bg _{\bullet,L_\bullet} \right]^{\ctrans} \bigg\},&
\end{aligned}
\end{equation}
then we obtain
\begin{equation}
\begin{aligned}
    \bar{\cov} _{\rmT} &= \frac{1}{L_\rmT}\bbE  \{\bH_{\rmT} \bH_{\rmT} ^{\ctrans} \} = \frac{ \cov _{\rmT,1} +\ldots, \cov _{\rmT,L_\rmT}} {L_\rmT},\\
    \bar{\cov} _{\rmR} & = \frac{1}{L_\rmR}\bbE  \{\bH_{\rmR} \bH_{\rmR} ^{\ctrans} \} = \frac {\cov _{\rmR,1} +\ldots, \cov_{\rmR,L_\rmR}}{L_\rmR}.
\end{aligned}
\end{equation}
Therefore, $\bH_{\rmR}$ and $\bH_{\rmT}$ can be approximated by the Karhunen-Loève transformation as $\bH_{\rmR} = \bar {\cov} ^{\half} _{\rmR} \bH_{w,\rmR}$ and $\bH_{\rmT} = \bar {\cov} ^{\half} _{\rmT} \bH_{w,\rmT}$, respectively, where $\bH_{w,\rmR}$ and $\bH_{w,\rmT}$ are complex Gaussian matrices with elements satisfying $\mathcal{CN}(0,1)$.

\section{Proof of Theorem \ref{theorem:capacitysparse}}
\label{proof:capacitysparseapp}
The ergodic SE under specular components dominant scenario is
\begin{equation}
	\begin{aligned}
		C_{\sparse} &= \bbE \left\{ \log_2 \det \left( \bI + \gamma \bH_{\sparse} \bH _{\sparse} ^{\ctrans} \right) \right\}\\
		& = \bbE \left\{ \log_2 \det \left( \bI + \gamma \tilde{\bA} \bPi_{\rmT}^{\ctrans} \bPi_{\rmT} \tilde{\bA}^{\ctrans} \bPi_\rmR^{\ctrans} \bPi_\rmR \right) \right\}.
	\end{aligned}
\end{equation}
Denoting the eigendecomposition of $\bPi_{\rmT}^{\ctrans} \bPi_{\rmT}$ and $\bPi_{\rmR}^{\ctrans} \bPi_\rmR$ as $\bPi_{\rmT}^{\ctrans} \bPi_{\rmT} = \bU_{\bPi_{\rmR}} \bLambda_{\bPi_{\rmR}} \bU^{\ctrans} _{\bPi_{\rmR}}$ and $\bPi_{\rmR}^{\ctrans} \bPi_{\rmR} = \bU_{\bPi_{\rmT}} \bLambda_{\bPi_{\rmT}} \bU^{\ctrans} _{\bPi_{\rmT}}$, respectively, the ergodic SE can be further represented as
\begin{equation}
\begin{aligned}
    C_{\sparse} & = \bbE \left\{ \log_2 \det \left( \bI + \gamma \blambda\left( \bLambda_{\bPi_{\rmR}} \tilde{\bA}^{\circ} \bLambda_{\bPi_{\rmT}} {\tilde{\bA}^{\circ}}{}^{\ctrans} \right) \right) \right\}\\
    &= \bbE \left\{ \log_2 \prod_{\ell=1}^{L} \left( 1 + \gamma \lambda_\ell \left(\bLambda_{\bPi_{\rmR}} \tilde{\bA}^{\circ} \bLambda_{\bPi_{\rmT}} {\tilde{\bA}^{\circ}}{}^{\ctrans} \right) \right) \right\},
\end{aligned}
\end{equation}
where $\tilde{\bA}^{\circ} = \bU^{\ctrans}_{\bPi_{\rmR}} \tilde{\bA} \bU_{\bPi_{\rmT}}$.
According to the majorization theory \cite{majorization}, the following weak majorization relationship is satisfied \cite{mmWcapacity}
\begin{equation}
    \blambda\left( \bLambda_{\bPi_{\rmR}} \tilde{\bA}^{\circ} \bLambda_{\bPi_{\rmT}} {\tilde{\bA}^{\circ}}{}^{\ctrans} \right)
    \! \prec_{w} \!
    \blambda(\bLambda_{\bPi_{\rmR}}) \odot \blambda(\bLambda_{\bPi_{\rmT}}) \odot \blambda(\tilde{\bA}^{\circ} {\tilde{\bA}^{\circ}} {}^{\ctrans}).
\end{equation}
For the accuracy of presentation, matrices $\blambda( \bLambda _{\bPi_{\rmR}})$, $ \blambda(\bLambda_{\bPi_{\rmT}})$, and $ \blambda( \tilde{\bA} ^{\circ} {\tilde{\bA} ^{\circ}} {}^{\ctrans})$ are rearranged as diagonal matrices with the dimension of $\rank(\bH_{\sparse}) \times \rank(\bH_{\sparse})$.
Denoting $L_{\sparse}=\rank(\bH_{\sparse})$, then we obtain the following approximation for the ergodic SE
\begin{equation}
\begin{aligned}
    C_{\sparse}^{\app} &\! = \! \bbE \left\{ \log_2 \det \left( \bI \! + \! \gamma \blambda(\bLambda_{\bPi_{\rmR}}) \! \odot \! \blambda(\bLambda_{\bPi_{\rmT}}) \! \odot \! \blambda(\tilde{\bA}^{\circ} {\tilde{\bA}^{\circ}} {}^{\ctrans}) \right) \right\}\\
    & \! = \! \bbE\left\{ \sum_{\ell=1}^{L_\sparse}\log_2 \left(1 \! + \! \gamma |\alpha_\ell|^2 \chi_{\rmR,\ell}\chi_{\rmT,\ell} \lambda _{\rmR,\ell} \lambda _{\rmT,\ell} |g_{\rmR,\ell} g^{*}_{\rmT,\ell}|^2 \right) \right\}.
\end{aligned}
\label{eq:capacitysparseapp1}
\end{equation}
Assmue $\dot{g}_\ell = |g_{\rmR,\ell}|^2 |g_{\rmT,\ell}|^2$, the PDF of $\dot{g}_\ell$ is $f_{\dot{G}_\ell}(g)=2K_0(2\sqrt{g})$ based on \textit{Lemma} \ref{lemma:exp} in Appendix \ref{sec:appexp}.
As a result, the approximation of ergodic SE in \eqref{eq:capacitysparseapp1} can be calculated as
\begin{equation}
\begin{aligned}
    C_{\sparse} ^{\app} &\!=\! \bbE_{\dot{g}_\ell}\left\{ \sum_{\ell=1}^{L_{\sparse}}\log_2 \left(1 \! + \! \gamma |\alpha_\ell|^2 \chi_{\rmR,\ell} \chi_{\rmT,\ell} \lambda_{\rmR,\ell}\lambda_{\rmT,\ell} \dot{g}_\ell \right) \right\}\\
    & \! = \! \sum_{\ell=1}^{L_{\sparse}} \int _{0} ^{\infty} 2\log_2 (1+\varpi _{\ell} \dot{g}) K_0(2\sqrt{\dot{g}}) \rmd \dot{g},
\end{aligned}
\label{eq:capacitysparseapp2}
\end{equation}
where $\varpi_\ell = \gamma |\alpha_\ell|^2 \chi_{\rmR,\ell}\chi_{\rmT,\ell} \lambda_{\rmR,\ell}\lambda_{\rmT,\ell}$.
Applying the Meijer-G function to represent the first term in the integral above \cite{MeijerG}, \eqref{eq:capacitysparseapp2} can be further written as
\begin{equation}
    \setlength\abovedisplayskip{5pt}
    \setlength\belowdisplayskip{5pt}
    C^{\app}_{\sparse} = \frac{1}{\ln 2}\sum _{\ell=1} ^{L_{\sparse}} G_{4,2} ^{1,4} \left( \varpi_\ell \Big| \begin{array}{cccc} 0&0&1&1 \\ 1&0&& \end{array}\right),
\end{equation}
with the aid of integral characteristics of $K_{\nu}(\cdot)$ and Meijer-G function \cite{TableInt}.

\section{Proof of Proposition \ref{prop:capacityrich}}
\label{proof:capacityrich}
According to \textit{Proposition} \ref{prop:richscatter}, under dense components scattering scenario, the channel can be approximated by
$\bH_{\RS} \approx \bar{\cov} ^{\half} _{\rmR} \bH_{w,\rmR} \bA \bH_{w,\rmT} ^ {\ctrans} \bar{\cov} ^{\halfH} _{\rmT}$.
We write $\bar{\cov}_{\rmR} = \bar{\bU}_{\rmR} \bar{\bLambda} _{\rmR} \bar{\bU} ^{\ctrans} _{\rmR}$, $\bar{\cov}_{\rmT} = \bar{\bU}_{\rmT} \bar{\bLambda} _{\rmT} \bar{\bU} ^{\ctrans} _{\rmT}$, and $\bA = \bU_{\sfC} \bLambda_{\sfC} \bU _{\sfC} ^{\ctrans}$, where $\bar{\bU}_{\rmR}$, $\bar{\bU} _{\rmT}$ and $\bU_{\sfC}$ are unitary matrices containing the respective eigenvectors, and $\bar{\bLambda}_{\rmR}$, $\bar{\bLambda}_{\rmT}$ and $\bLambda _{\sfC}$ are diagonal matrices with diagonal elements pertaining to the respective descending ordered eigenvalues.
Then the ergodic capacity is further represented as
\begin{equation}
\begin{aligned}
    C_{\RS} = \bbE \left\{ \log_2 \det \left( \bI +  \gamma \bar{\bLambda}_{\rmR} \tilde{\bH}_{w,\rmR} \bLambda _{\sfC} \tilde{\bH} ^{\ctrans} _{w,\rmT} \times \right. \right.&\\
    \left. \left. \bar{\bLambda}_{\rmT} \tilde{\bH} _{w,\rmT} \bLambda ^{\ctrans} _{\sfC} \tilde{\bH} ^{\ctrans} _{w,\rmR} \right) \right\}&,
\end{aligned}
\label{eq:capacityrich1}
\end{equation}
where $\tilde{\bH}_{w,\rmR} = \bar{\bU} ^{\ctrans} _{\rmR} \bH_{w,\rmR} \bU_{\sfC}$ and $\tilde{\bH}_{w,\rmT} = \bar{\bU} ^{\ctrans} _{\rmT} \bH_{w,\rmT} \bU_{\sfC}$ are still Gaussian matrix variates with i.i.d. $\mathcal{CN}(0,1)$ elements.
Then, applying the following determinant expansion \cite{detexpansion}
\begin{equation}
    \det(\bI_{n} + \bM) = \sum _{k=0} ^{n} \sum_{\hat{\alpha} _k} \det(\bM)_{\hat{\alpha}_k} ^{\hat{\alpha}_k},
\end{equation}
and the Cauchy-Binet theorem \cite{majorization}, the determinant term in \eqref{eq:capacityrich1} can be denoted by exchanging the order of the product of determinants as
\begin{equation}
\begin{aligned}
C_{\RS} = \bbE \bigg\{ \log_2 \bigg( \sum_{k=0} ^ {r}\gamma^k \sum_{\hat{\alpha}_1}\dots \sum_{\hat{\alpha}_8} \det(\bar{\bLambda}_{\rmR}) ^{\hat{\alpha}_1} _{\hat{\alpha}_1} &\\
\times \det({\bLambda}_{\sfC})^{\hat{\alpha}_2} _{\hat{\alpha}_3} \times \det(\bLambda_{\sfC} ^{\ctrans}) ^{\hat{\alpha}_6} _{\hat{\alpha}_7} \times \det(\bar{\bLambda}_{\rmT}) ^{\hat{\alpha}_4} _{\hat{\alpha}_5} & \\
\times \det(\tilde{\bH}_{w,\rmR})  ^{\hat{\alpha}_1} _{\hat{\alpha}_2} \times \det(\tilde{\bH}^{\ctrans} _{w,\rmR}) ^{\hat{\alpha}_7} _{\hat{\alpha}_8} &\\
\times \det(\tilde{\bH}^{\ctrans} _{w,\rmT}) ^{\hat{\alpha}_3} _{\hat{\alpha}_4} \times \det(\tilde{\bH} _{w,\rmT}) ^{\hat{\alpha}_5} _{\hat{\alpha}_6} & \bigg) \bigg\}.
\end{aligned}
\end{equation}
Assuming the $\bA$ is a positive semi-definite matrix, the Jensen's inequality $\bbE \{ \log \det (\bM) \} \leq \log \bbE \{ \det (\bM) \}$ can be applied. Combined with the following properties in \cite{keyhole}
\begin{equation}
\begin{aligned}
    \bbE \left\{ \det(\bH_{w})^{\hat{\alpha}_k ^p} _{\hat{\alpha}_k ^q} \right.& \left. \det(\bH_{w}) ^{\hat{\alpha}_k ^s} _{\hat{\alpha}_k ^t}  \right\}\\
    &= \begin{cases}
k!,&{\text{if}} \ \hat{\alpha}_k ^p=\hat{\alpha}_k ^t,\hat{\alpha}_k ^q=\hat{\alpha}_k ^s,\\
{0,}&{\text{else,}}
\end{cases}
\end{aligned}
\end{equation}
where $\bH_{w}$ is a matrix with entries satisfy i.i.d. $\mathcal{CN}(0,1)$, the expression in \eqref{eq:capacityrich} is obtained.

%\vspace{-0.8cm}
\section{Proof of Theorem \ref{theorem:outsparse}}
\label{proof:outsparse}
According to the majorization relationship of the eigenvalues of $\bH _{\sparse}^{\ctrans} \bH_{\sparse}$ presented in Appendix \ref{proof:capacitysparseapp}, the $\ell$-th largest eigenvalue of $\bH _{\sparse} ^{\ctrans} \bH_{\sparse}$ satisfies
\begin{equation}
    \lambda_{\ell} (\bH _{\sparse} ^{\ctrans} \bH_{\sparse}) \approx \lambda_\ell(\bLambda_{\bPi_{\rmR}}) \lambda_\ell(\bLambda_{\bPi_{\rmT}}) \lambda_\ell(\tilde{\bA}^{\circ}{} ^{\ctrans} {\tilde{\bA}^{\circ}}) = \bar{\lambda} ^{\sparse} _{\ell},
\end{equation}
where $\bar{\lambda} ^{\sparse} _{\ell} = |\alpha_\ell|^2 \chi_{\rmR,\ell} \chi_{\rmT,\ell} \lambda_{\rmR,\ell} \lambda_{\rmT,\ell}  |\dot{g}_{\ell}|^2 $.
Then $P_{\out,\sparse}$ can be further written as \cite{order}
\begin{equation}
\begin{aligned}
    P_{\out,\sparse} &= \Prob \left( \lambda_{\max}(\bH _{\sparse} ^{\ctrans} \bH_{\sparse}) \leq \frac{\gamma_{\threshold}}{\bar{\gamma}}\right) \\
    &= \Prob \left(\lambda_{\ell}(\bH _{\sparse} ^{\ctrans} \bH_{\sparse}) \leq \frac{\gamma_{\threshold}}{\bar{\gamma}} \right),\text{for}\ \ell=1,\ldots,L_{\sparse}\\
    & \overset{(a)}{=} F_{\lambda_{1}(\bH _{\sparse} ^{\ctrans} \bH_{\sparse})}\left(\frac{\gamma_{\threshold}}{\bar{\gamma}}\right) \dots \cdot F_{\lambda_{L _\sparse}(\bH _{\sparse} ^{\ctrans} \bH_{\sparse})} \left(\frac{\gamma_{\threshold}}{\bar{\gamma}}\right)\\
    & \approx F_{\bar{\lambda} ^{\sparse}_{1}} \left(\frac{\gamma_{\threshold}}{\bar{\gamma}}\right) \dots \cdot F_{\bar{\lambda} ^{\sparse} _{L}}\left(\frac{\gamma_{\threshold}}{\bar{\gamma}}\right),
\end{aligned}
\end{equation}
where $(a)$ is because $\lambda_{\ell}(\bH _{\sparse} ^{\ctrans} \bH_{\sparse})$ are independent for $\ell=1,\ldots,L _{\sparse}$.
Then combined with the CDF in Appendix \ref{sec:appexp} and the proof is completed.
}

\vfill


\begin{thebibliography}{1}
\bibliographystyle{IEEEtran}

\bibitem{MatthaiouCM}
M. Matthaiou, \textit{et al}., ``The road to 6G: Ten physical layer challenges for communications engineers," \textit{IEEE Commun. Mag.}, vol. 59, no. 1, pp. 64-69, Jan. 2021.

\bibitem{HanIOTJ}
Y. Han, S. Jin, M. Matthaiou, T. Q. S. Quek, and C. -K. Wen, ``Toward extra large-scale MIMO: New channel properties and low-cost designs," \textit{IEEE Internet Things J.}, vol. 10, no. 16, pp. 14569-14594, Aug. 2023.

\bibitem{nokia}
Nokia, ``6G mid-band spectrum technology explained," Available: https://www.nokia.com/about-us/newsroom/articles/6g-mid-band-spectrum-technology-explained/.

\bibitem{WRC23}
ITU, ``World Radiocommunication Conference 2023 (WRC-23)," Available: https://www.itu.int/dms\_pub/itu-r/opb/act/R-ACT-WRC.15-2023-PDF-E.pdf, Dec. 2023.

\bibitem{U6G}
M. Ghosh, ``Evolution of sharing in 6 GHz," \textit{IEEE Wireless Commun.}, vol. 30, no. 5, pp. 4-5, Oct. 2023.

\bibitem{U6G1}
N. Li, C. Guo, and D. Wang, ``Considerations on 6 GHz spectrum for 5G-Advanced and 6G," \textit{IEEE Commun. Standards Mag.}, vol. 5, no. 3, pp. 5-7, Sept. 2021.

\bibitem{spectrum}
S. Kang, G. Geraci, M. Mezzavilla, and S. Rangan, ``Terrestrial-satellite spectrum sharing in the upper mid-band with interference nulling," \textit{arXiv:2311.12965}, Nov. 2023.

\bibitem{RT}
S. Kang \textit{et al.}, ``Cellular wireless networks in the upper mid-band," \textit{IEEE Open J. Commun. Soc.}, vol. 5, pp. 2058-2075, 2024.

\bibitem{MiaoJSAC}
H. Miao et al., ``Sub-6 GHz to mmWave for 5G-advanced and beyond: Channel measurements, characteristics and impact on system performance," \textit{IEEE J. Sel. Areas Commun.}, vol. 41, no. 6, pp. 1945-1960, Jun. 2023.

\bibitem{TSR}
D. Shakya \textit{et al}., ``Propagation measurements and channel models in indoor environment at 6.75 GHz FR1(C) and 16.95 GHz FR3 upper-mid band spectrum for 5G and 6G", \textit{arXiv:2405.01358}, May 2024.

\bibitem{DMC1}
A. Richter, ``Estimation of radio channel parameters: Models and algorithms," Ph.D. dissertation, ISLE Blacksburg, VA, USA, 2005.

\bibitem{DMC2}
S. Jiang, W. Wang, Y. Miao, W. Fan, and A. F. Molisch, ``A survey of dense multipath and its impact on wireless systems," \textit{IEEE Open J. Antennas Propag.}, vol. 3, pp. 435-460, Apr. 2022.

\bibitem{TianWCL}
J. Tian, Y. Han, S. Jin, and M. Matthaiou, "Low-overhead localization and VR identification for subarray-based ELAA systems," \textit{IEEE Wireless Commun. Lett.}, vol. 12, no. 5, pp. 784-788, May 2023.

\bibitem{LuICC}
H. Lu and Y. Zeng, ``How does performance scale with antenna number for extremely large-scale MIMO?" in \textit{Proc. IEEE ICC}, 2021, pp. 1-6.

\bibitem{YangTVT}
X. Yang, F. Cao, M. Matthaiou, and S. Jin, ``On the uplink transmission of extra-large scale massive MIMO systems," \textit{IEEE Trans. Veh. Technol.}, vol. 69, no. 12, pp. 15229-15243, Dec. 2020.

\bibitem{LDMA}
Z. Wu and L. Dai, ``Multiple access for near-field communications: SDMA or LDMA?" \textit{IEEE J. Sel. Areas Commun.}, vol. 41, no. 6, pp. 1918-1935, Jun. 2023.

\bibitem{NFCor}
Z. Dong and Y. Zeng, ``Near-field spatial correlation for extremely large-scale array communications," \textit{IEEE Commun. Lett.}, vol. 26, no. 7, pp. 1534-1538, Jul. 2022.

\bibitem{XLmeasure}
H. Miao, \textit{et al}., ``Empirical studies of propagation characteristics and modeling based on XL-MIMO channel measurement: From far-field to near-field," \textit{arXiv: 2404.17270}, Apr. 2024.

\bibitem{nonstationary}
E. D. Carvalho, A. Ali, A. Amiri, M. Angjelichinoski, and R. W. Heath, "Non-stationarities in extra-large-scale massive MIMO," \textit{IEEE Wireless Commun.}, vol. 27, no. 4, pp. 74-80, Aug. 2020.

\bibitem{U6Gsparse}
X. Liu \textit{et al}., ``Channel sparsity variation and model-based analysis on 6, 26, and 105 GHz measurements," \textit{IEEE Trans. Veh. Technol.}, vol. 73, no. 7, pp. 9387-9397, Jul. 2024.

\bibitem{channelsurvey}
R. Feng, C. -X. Wang, J. Huang, X. Gao, S. Salous, and H. Haas, ``Classification and comparison of massive MIMO propagation channel models," \textit{IEEE Internet Things J.}, vol. 9, no. 23, pp. 23452-23471, Dec. 2022.

\bibitem{finite}
A. G. Burr, ``Capacity bounds and estimates for the finite scatterers MIMO wireless channel," \textit{IEEE J. Sel. Areas Commun.}, vol. 21, no. 5, pp. 812-818, Jun. 2003.

\bibitem{VCR}
A. M. Sayeed, ``Deconstructing multiantenna fading channels," \textit{IEEE Trans. Signal Proces.}, vol. 50, no. 10, pp. 2563-2579, Oct. 2002.

\bibitem{Kronecker}
Da-Shan Shiu, G. J. Foschini, M. J. Gans, and J. M. Kahn, ``Fading correlation and its effect on the capacity of multielement antenna systems," \textit{IEEE Trans. Commun.}, vol. 48, no. 3, pp. 502-513, Mar. 2000.

\bibitem{double}
D. Gesbert, H. Bolcskei, D. A. Gore, and A. J. Paulraj, ``Outdoor MIMO wireless channels: Models and performance prediction," \textit{IEEE Trans. Commun.}, vol. 50, no. 12, pp. 1926-1934, Dec. 2002.

\bibitem{Weichselberger}
W. Weichselberger, M. Herdin, H. Ozcelik, and E. Bonek, ``A stochastic MIMO channel model with joint correlation of both link ends," \textit{IEEE Trans. Wireless Commun.}, vol. 5, no. 1, pp. 90-100, Jan. 2006.

\bibitem{CJE}
J. Tian, Y. Han, S. Jin, J. Zhang, and J. Wang, ``Analytical channel modeling: From MIMO to extra large-scale MIMO,” \textit{Chin. J. Electron.}, early access, doi: 10.23919/cje.2023.00.418.

\bibitem{38901}
3GPP, ``Study on channel model for frequencies from 0.5 to 100 GHz (Release 18),” 3GPP TR 38.901, \textit{Technical Report}, Mar. 2024.

\bibitem{GBSM1}
S. Wu, C. -X. Wang, e. -H. M. Aggoune, M. M. Alwakeel, and Y. He, ``A non-stationary 3-D wideband twin-cluster model for 5G massive MIMO channels," \textit{IEEE J. Sel. Areas Commun.}, vol. 32, no. 6, pp. 1207-1218, Jun. 2014.

\bibitem{25996}
3GPP, ``Spatial channel model for Multiple Input Multiple Output (MIMO) simulations (Release 17)," 3GPP TR 25.996, \textit{Technical Report}, Mar. 2022.

\bibitem{NFLoS}
Y. Liu, Z. Wang, J. Xu, C. Ouyang, X. Mu, and R. Schober, ``Near-field communications: A tutorial review," \textit{IEEE Open J. Commun. Soc.}, vol. 4, pp. 1999-2049, 2023.

\bibitem{Fourier}
A. Pizzo, L. Sanguinetti, and T. L. Marzetta, "Fourier plane-wave series expansion for holographic MIMO communications," \textit{IEEE Trans. Wireless Commun.}, vol. 21, no. 9, pp. 6890-6905, Sept. 2022.

\bibitem{Fresnel}
J. Sherman, ``Properties of focused apertures in the fresnel region," IRE Transactions on Antennas and Propagation, vol. 10, no. 4, pp. 399-408, Jul. 1962.

\bibitem{NFUPA}
B. Friedlander, ``Localization of signals in the near-field of an antenna array," \textit{IEEE Trans. Signal Proces.}, vol. 67, no. 15, pp. 3885-3893, Aug. 2019.

\bibitem{KL}
L. Sanguinetti, et al., ``Toward massive MIMO 2.0: Understanding spatial correlation, interference suppression, and pilot contamination," \textit{IEEE Trans. Commun.}, vol. 68, no. 1, pp. 232-257, Jan. 2020.

\bibitem{onering}
A. Abdi and M. Kaveh, ``A space-time correlation model for multielement antenna systems in mobile fading channels," \textit{IEEE J. Sel. Areas Commun.}, vol. 20, no. 3, pp. 550-560, Apr. 2002.

\bibitem{dist1}
Z. Ma \textit{et al}., ``Impact of UAV rotation on MIMO channel characterization for air-to-ground communication systems," \textit{IEEE Trans. Veh. Technol.}, vol. 69, no. 11, pp. 12418-12431, Nov. 2020.

\bibitem{dist2}
W. Tan and S. Ma, ``Antenna array topologies for mmWave massive MIMO systems: Spectral efficiency analysis," \textit{IEEE Trans. Veh. Technol.}, vol. 71, no. 12, pp. 12901-12915, Dec. 2022.

\bibitem{multikeyhole1}
G. Levin and S. Loyka, ``From multi-keyholes to measure of correlation and power imbalance in MIMO channels: Outage capacity analysis," \textit{IEEE Trans. Inform. Theory}, vol. 57, no. 6, pp. 3515-3529, Jun. 2011.

\bibitem{ToepDFT}
R. Gray, ``On the asymptotic eigenvalue distribution of Toeplitz matrices," \textit{IEEE Trans. Inform. Theory}, vol. 18, no. 6, pp. 725-730, Nov. 1972.

\bibitem{FFGao}
H. Xie, F. Gao, S. Jin, J. Fang, and Y. -C. Liang, ``Channel estimation for TDD/FDD massive MIMO systems with channel covariance computing," \textit{IEEE Trans. Wireless Commun.}, vol. 17, no. 6, pp. 4206-4218, Jun. 2018.

\bibitem{SMBF}
A. Forenza, M. R. McKay, A. Pandharipande, R. W. Heath, and I. B. Collings, ``Adaptive MIMO transmission for exploiting the capacity of spatially correlated channels," \textit{IEEE Trans. Veh. Technol.}, vol. 56, no. 2, pp. 619-630, Mar. 2007.

\bibitem{NFfinite}
J. Tian, Y. Han, S. Jin, X. Li, J. Zhang, and M. Matthaiou, ``Near-field channel reconstruction in sensing RIS-assisted wireless communication systems," \textit{IEEE Trans. Wireless Commun.}, early access, doi: 10.1109/TWC.2024.3389026.

\bibitem{keyhole}
H. Shin and J. H. Lee, ``Capacity of multiple-antenna fading channels: Spatial fading correlation, double scattering, and keyhole," \textit{IEEE Trans. Inform. Theory}, vol. 49, no. 10, pp. 2636-2647, Oct. 2003.

\bibitem{McKay}
M. R. McKay, A. J. Grant, and I. B. Collings, ``Performance analysis of MIMO-MRC in double-correlated Rayleigh environments," \textit{IEEE Trans. Commun.}, vol. 55, no. 3, pp. 497-507, Mar. 2007.

\bibitem{quadratic}
T. Ratnarajah and R. Vaillancourt, ``Quadratic forms on complex random matrices and multiple-antenna systems," \textit{IEEE Trans. Inform. Theory}, vol. 51, no. 8, pp. 2976-2984, Aug. 2005.

\bibitem{38104}
3GPP, ``Base Station (BS) radio transmission and reception (Release 18)," 3GPP TS 38.104, \textit{Technical Report}, Mar. 2024.

\bibitem{TableInt}
I. Gradshteyn and I. Ryzhik, \textit{Table of Integrals, Series, and Products}. Academic Press, 1980.

\bibitem{matvariate}
A. K. Gupta and D. K. Nagar, \textit{Matrix Variate Distributions}, vol. 104. Boca Raton, FL, USA: Chapman \& Hall, 2018.

\bibitem{JHWang}
J. Wang, H. Wang, Y. Han, S. Jin, and X. Li, ``Joint transmit beamforming and phase shift design for reconfigurable intelligent surface assisted MIMO systems," \textit{IEEE Trans. Cogn. Commun. Netw.}, vol. 7, no. 2, pp. 354-368, Jun. 2021.

\bibitem{majorization}
A. W. Marshall and I. Olkin, \textit{Inequalities: Theory of Majorization and Its Applications}. New York, NY, USA: Academic, 1979.

\bibitem{mmWcapacity}
X. Yang, X. Li, S. Zhang, and S. Jin, ``On the ergodic capacity of mmWave systems under finite-dimensional channels," \textit{IEEE Trans. Wireless Commun.}, vol. 18, no. 11, pp. 5440-5453, Nov. 2019.

\bibitem{MeijerG}
A. P. Prudnikov, Y. A. Brychkov, and O. I. Marichev, \textit{Integrals and Series}. New York: Gordon and Breach, 1990, vol. 3, More Special Functions.

\bibitem{detexpansion}
Q. T. Zhang, X. W. Cui, and X. M. Li, ``Very tight capacity bounds for MIMO-correlated Rayleigh-fading channels," \textit{IEEE Trans. Wireless Commun.}, vol. 4, no. 2, pp. 681-688, Mar. 2005.

\bibitem{order}
H. A. David and H. N. Nagaraja, \textit{Order Statistics}, 3rd ed. Hoboken, NJ, USA: Wiley, 2003.


\end{thebibliography}
\end{document}